\documentclass[a4paper,11pt]{article}
\usepackage{jheppub}
\usepackage[T1]{fontenc}
\usepackage{graphicx,subcaption}
\usepackage{braket}
\usepackage{bm, comment}
\usepackage{xcolor}

\definecolor{purple}{rgb}{0.3,0,0.7}

\usepackage{acro}
\DeclareAcronym{gr}{
    short=GR ,
    long=general relativity
}
\DeclareAcronym{uv}{
    short=UV ,
    long=ultraviolet
}
\DeclareAcronym{ir}{
    short=IR ,
    long=infrared
}
\DeclareAcronym{rg}{
    short=RG ,
    long=renormalization group
}
\DeclareAcronym{qg}{
    short=QG ,
    long=quantum gravity
}
\DeclareAcronym{kk}{
    short=KK ,
    long=Kaluza-Klein
}
\DeclareAcronym{qft}{
    short=QFT ,
    long=quantum field theory 
}

\DeclareAcronym{flrw}{
    short=FLRW ,
    long=Friedmann-Lema\^{i}tre-Robertson-Walker
}
\DeclareAcronym{sm}{
    short=SM ,
    long=standard model
}
\allowdisplaybreaks[1]

\title{\bf Casimir-Induced Quintessence in Dark Dimension}

\author{\large Tomoki Katayama${}^{a,b}$, Hiroki Matsui${}^{c,d}$, Yuri Michinobu${}^{d}$, Fumiya Okamatsu${}^{c}$, Yutaka Sakamura${}^{a,b}$, Takahiro Terada${}^{e}$}

\emailAdd{tomokika@post.kek.jp} 
\emailAdd{hiroki.matsui@yukawa.kyoto-u.ac.jp}
\emailAdd{yuri.michinobu@yukawa.kyoto-u.ac.jp} 
\emailAdd{okamatsu.fumiya@nihon-u.ac.jp} 
\emailAdd{sakamura@post.kek.jp}
\emailAdd{terada@eken.phys.nagoya-u.ac.jp}

\affiliation{
${}^a$Institute of Particle and Nuclear Studies, High Energy Accelerator Research Organization (KEK), Oho 1-1, Tsukuba 305-0801, Japan \medskip\\
${}^b$The Graduate University for Advanced Studies (SOKENDAI), Tsukuba 305-0801, Japan \medskip\\ 
${}^c$Department of Physics, College of Humanities and Sciences, Nihon University, Sakurajosui, Tokyo 156-8550, Japan \medskip\\
${}^d$Center for Gravitational Physics and Quantum Information, Yukawa Institute for Theoretical Physics, Kyoto University, Kitashirakawa Oiwakecho, Sakyo-ku, Kyoto 606-8502, Japan \medskip \\
${}^e$Kobayashi-Maskawa Institute for the Origin of Particles and the Universe, 
Nagoya University, Nagoya 464-8602, Japan}

\abstract{We investigate a concrete realization of the Dark Dimension scenario, where a single large extra dimension is set at sub-millimeter scales. In this framework, the Casimir energy of bulk fields accounts for the observed dark energy. Working in a 5-dimensional setup with the Standard Model confined to a 4-dimensional brane, we derive the effective action for the radion. We demonstrate that a minimal model comprising only gravity and three right-handed bulk neutrinos typically yields a negative radion potential. To realize a positive vacuum energy, we consider some extensions with additional bulk degrees of freedom. These extensions generate a sufficiently flat positive potential that allows the radion to behave as a quintessence field, evolving slowly at the sub-eV scale. Finally, we analyze the evolution of the dark-energy equation-of-state parameter and show that our model is consistent with recent DESI BAO measurements, including the distance ratios $D_H/r_d$ and $D_M/r_d$.}
\begin{document}
\maketitle
\flushbottom

\section{Introduction}
\label{sec:introduction}

Observations of Type Ia supernovae~\cite{Riess:1998cb,Perlmutter:1998np}, the cosmic microwave background (CMB)~\cite{Spergel:2003cb,Komatsu:2010fb}, and baryon acoustic oscillations (BAO)~\cite{Eisenstein:1997ik,SDSS:2005xqv,Bassett:2009mm} have established that the expansion of the Universe is currently accelerating. This late-time acceleration is usually attributed to a dark-energy component that accounts for roughly $70\%$ of the present energy density. The simplest candidate is the cosmological constant, but this raises serious theoretical puzzles. The observed vacuum-energy scale is tiny compared to any known \ac{uv} scale~\cite{Weinberg:1988cp}, and the so-called coincidence problem, why the dark-energy and matter densities are comparable precisely today, remains unexplained~\cite{Zlatev:1998tr,Steinhardt:1999nw}. These tensions motivate scenarios in which the dark energy is dynamical and potentially linked to physics beyond the \ac{sm}~\cite{Copeland:2006wr}.

String theory, as a leading candidate for a consistent theory of quantum gravity, naturally introduces extra dimensions~\cite{Green:1987sp,Polchinski:1998rq}. In traditional models, these dimensions are taken to be of the Planck length order, but large or even macroscopic extra dimensions are possible in brane-world scenarios~\cite{Rubakov:1983bz,Gogberashvili:1998vx,Gogberashvili:1998iu,Gogberashvili:1999tb,Randall:1999ee,Randall:1999vf}. In such frameworks, the \ac{sm} fields are localized in a 4-dimensional brane, whereas gravity propagates in the higher-dimensional bulk. The compactification scale, set by the inverse size of the extra dimensions, then plays a central role both in determining the effective 4-dimensional Planck scale and in controlling quantum effects such as Casimir energy.

Quantum fluctuations in compact extra dimensions generate a Casimir energy that depends sensitively on the compactification radius~$R$~\cite{Appelquist:1983vs}. The resulting Casimir potential can stabilize moduli and fix the size of the extra dimensions~\cite{Garriga:2000jb,Toms:2000bh,Goldberger:2000dv,Brevik:2000vt,Greene:2007xu}, potentially in conjunction with other ingredients such as fluxes and non-perturbative effects~\cite{Kachru:2003aw,Balasubramanian:2005zx}. 

More recently, the Dark Dimension scenario~\cite{Montero:2022prj} has been proposed in the context of the swampland program. This proposal relies on the (Anti-)de Sitter ((A)dS) distance conjecture~\cite{Lust:2019zwm} and the emergent string conjecture~\cite{Lee:2019xtm, Lee:2019wij}. The Swampland Distance Conjecture~\cite{Ooguri:2006in} tells us that an infinite tower of states becomes light in asymptotic regions in the moduli space.  Such tower states are expected also in the limit of a small absolute value of the cosmological constant according to the AdS distance conjecture and its generalization to de Sitter spacetime. On the other hand, the emergent string conjecture tells us that the tower states are either the fundamental string states (weak-coupling limit) or the \ac{kk} states (decompactification limit).  In the Dark Dimension scenario, one interprets the smallness of the observed cosmological constant as a presence of light tower states whose mass scale is the sub-eV scale.  Such a low string scale is completely excluded experimentally, so one is led to conclude there must be a micron-sized extra dimension ($R \sim \Lambda_{\rm c.c.}^{-1/4} \sim \mu\text{m}$) and the associated \ac{kk} modes. 
In this picture, the size of the extra dimension is closely related to the dark-energy scale. The \ac{kk} Casimir energy of light bulk fields provides the dominant contribution to the present vacuum energy, in a manner consistent with the swampland conjectures.

Motivated by this idea, we investigate a concrete realization of the Dark Dimension scenario and its cosmological phenomenology. We consider a 5-dimensional gravitational theory with a single compact extra dimension. The \ac{sm} fields are localized on a 4-dimensional brane, while gravity and additional gauge fields propagate in the bulk. In particular, the bulk matter sector includes three generations of right-handed (sterile) neutrinos. This setup is well motivated if the compactification scale is related to the neutrino mass scale, since small neutrino masses can naturally arise from suppressed effective 4-dimensional neutrino Yukawa couplings.

We compute the resulting Casimir energy, study its role in determining the dynamics of the extra dimension, and analyze the effective 4-dimensional dynamics of the radion field as a candidate for dynamical dark energy. Several analyses of Dark Dimension models and related constructions have recently appeared in the literature, with emphasis on effective field-theory considerations and model building~\cite{Burgess:2023pnk,Branchina:2023ogv,Anchordoqui:2023wkm,Anchordoqui:2023laz,Cui:2023wzo,Heckman:2024trz}. In contrast, our focus is on the explicit realization of the Dark Dimension with the Casimir-induced radion potential, and the resulting cosmological evolution, including the dark-energy equation of state and constraints from cosmological observations.%
\footnote{Within a more general framework of the Dark Dimension, the possibility of dynamical dark energy and its consistency with observations such as DESI were discussed in Ref.~\cite{Bedroya:2025fwh}. That analysis is complementary to the present work, since we provide an explicit realization of a Casimir-induced radion potential within the Dark-Dimension scenario and study its cosmological implications.}
We show that, in the minimal setup containing only 5-dimensional gravity and three generations of bulk neutrinos, the effective potential generically fails to provide a viable positive region, implying that additional bulk contributions are required to realize a positive dark-energy density. With such extensions, the radion can be set at a sub-eV compactification scale and gives rise to a quintessence dark-energy component with a mildly evolving equation-of-state parameter.
This matches the recent Dark Energy Spectroscopic Instrument (DESI) constraints~\cite{DESI:2024mwx, DESI:2024hhd, DESI:2025zgx, DESI:2025fii}.

The remainder of this paper is organized as follows. In Section~\ref{sec:higher-dimensional-gravity}, we define the 5-dimensional gravitational framework and perform the dimensional reduction. Sec.~\ref{sec:casimir-stabilization} details the calculation of the Casimir energy and discusses the conditions for a realization of the Dark Dimension scenario. In Sec.~\ref{sec:dynamical-dark-energy}, we derive the 4-dimensional effective action, analyze the radion-quintessence dynamics, and compare the predictions of our model with 
the current cosmological observations, including DESI data. Finally, our conclusions are summarized in Sec.~\ref{sec:conclusion}.

\section{Setup}
\label{sec:higher-dimensional-gravity}

As a concrete realization of the Dark Dimension scenario, we consider a $5$-dimensional gravitational theory in which the \ac{sm} fields and dark matter fields are localized on $4$-dimensional branes, while gravity and some additional matter fields propagate in the full $5$-dimensional bulk. In this framework, the single extra dimension plays the role of the Dark Dimension, and its dynamics and stabilization are central to the cosmological evolution.

We start from the $5$-dimensional Einstein-Hilbert action with a bulk cosmological constant $\Lambda_{5}$ and matter fields:
\begin{equation}
    S = \frac{1}{16\pi G_{5}} \int d^{5}x \sqrt{-g_{5}} \left( {\cal R}_{5} - 2\Lambda_{5} \right) 
    + S_{\text{brane}}^{\text{SM}} + S_{\text{brane}}^{\text{DM}} \,.
    \label{eq:HDaction}
\end{equation}
Here, $g_{5}$ is the determinant of the $5$-dimensional metric, ${\cal R}_{5}$ is the corresponding Ricci scalar, and the $5$-dimensional Newton constant $G_{5}$ is related to the fundamental Planck mass $M_{5}$ via $G_{5} = M_{5}^{-3}$. The $S_{\text{brane}}^{\text{SM}}$ and $S_{\text{brane}}^{\text{DM}}$ represent the actions of the \ac{sm} fields and the dark matter sector, which may in general be localized on distinct 4-dimensional branes.
For definiteness, the \ac{sm} action on the brane is written as
\begin{equation}
    S_{\text{brane}}^{\text{SM}}
    = \int d^{4}x \sqrt{-g^{\rm ind}}\,
    \mathcal{L}_{\text{SM}}
    \bigl(\psi, g^{\rm ind}_{\mu\nu}\bigr) \,,
    \label{eq:brane-action}
\end{equation}
where $g^{\rm ind}_{\mu\nu}$ is the metric induced on the brane and $\psi$ denotes the \ac{sm} fields. An analogous expression applies to 
$S_{\text{brane}}^{\text{DM}}$ for the dark matter sector.

We assume that the bulk geometry is flat, and 
adopt the diagonal metric ansatz
\begin{equation}
  ds^2 = g_{\mu\nu}(x) \, dx^{\mu} dx^{\nu} + b^2(x)\,dy^2,
  \label{eq:metric-ansatz}
\end{equation}
where $g_{\mu\nu}(x)$ 
is the metric of the 4-dimensional spacetime, and $b(x)$ is the radion field that reduces to the scale factor of the extra dimension, $b(t)$, for a homogeneous cosmological background. The coordinate volume of the compact dimension is ${\cal V}\equiv 2\pi R_{0}$, where $R_0$ is the (comoving) radius of the extra dimension.

Using the standard expressions for the higher-dimensional Ricci scalar evaluated on the ansatz~\eqref{eq:metric-ansatz}, and integrating over the compact coordinate, we obtain the Jordan-frame 4-dimensional action~\cite{Carroll:2001ih}
\begin{align}
\begin{split}
S&=\frac{M_{5}^{3}{\cal V}}{16\pi}\!\int d^{4}x\sqrt{-g}
\bigl[b{\cal R}_4 -2g^{\mu\nu}\nabla_\mu \nabla_\nu b 
   -2b\Lambda_5\bigr] + S_{\text{brane}}^{\text{SM}} + S_{\text{brane}}^{\text{DM}}   \,,
\label{eq:Jordan-frame}
\end{split}
\end{align}
where ${\cal R}_4$ is the 4-dimensional Ricci scalar. The 4-dimensional Planck mass is defined as 
\begin{equation}\label{eq:4-Planck-mass}
M_{\rm{Pl}}^{2} \equiv {\cal V}M_{5}^{3}\,,
\end{equation}
in agreement with the usual relation between the higher-dimensional and 4-dimensional Planck scales. Later, we introduce the effective cosmological constant $\Lambda_{\text{eff}}(b)$ which incorporates the Casimir energy contribution, 
\begin{equation}
\Lambda_{\text{eff}}(b) = \Lambda_5 + \frac{8\pi {\cal V}}{M_{\text{Pl}}^2} \rho_{\text{C}}(b).
\label{eq:effective-cosmo-constant}
\end{equation}

The relation~\eqref{eq:4-Planck-mass} connects the fundamental $5$-dimensional Planck scale $M_{5}$ to the observed $4$-dimensional Planck scale $M_{\rm{Pl}}$ via the volume of the extra dimension. The Dark Dimension scenario assumes that the physical size of the extra dimension, $bR_0$, is tied to the dark-energy scale and hence lies in the sub-millimetre range. 
This leads to a fundamental scale of order $M_{5} \sim 10^9\, {\rm GeV}$. This value lies in an intermediate range, far below the $4$-dimensional Planck scale but well above the electroweak scale. 

Here, let us give a brief comment on a case where there are two large extra 
dimensions compactified on a torus $T^2$. 
In such a case, the same requirement on the dark-energy scale would imply a $6$-dimensional Planck mass of order $M_{6} \sim {\cal O}({\rm TeV})$, suggesting an intriguing connection between the hierarchy problem and the dark energy problem. However, this possibility is under strong pressure from collider experiments, in particular from the absence of microscopic black-hole signatures and other characteristic signals of TeV-scale gravity at the LHC~\cite{Banks:1999eg}. Additional constraints arise from astrophysical considerations, especially the cooling of neutron stars~\cite{Cullen:1999hc,Barger:1999jf,Hanhart:2000er,Hanhart:2001fx,Hannestad:2003yd,Anchordoqui:2025nmb,Hardy:2025ajb}.

Neutron stars formed immediately after a supernova explosion are extremely hot, and at high temperatures we expect copious production of \ac{kk} gravitons (and other bulk modes, if present). Some of these modes remain gravitationally bound to the neutron star and subsequently decay into photon pairs, part of whose energy is reabsorbed by the star. This delayed reheating slows down the cooling of the neutron stars. Observations of sufficiently cold neutron stars, therefore, constrain the total emissivity into bulk modes. For two extra dimensions, the resulting bound roughly implies $b R_0 \lesssim 10^{-4}\, \mu\textrm{m}$, which is incompatible with the micron-scale radius suggested by the dark-energy scale and disfavors this Dark Dimension scenario. By contrast, for a single extra dimension, the constraints from both tabletop gravity experiments and neutron-star cooling are significantly weaker. Neutron-star cooling bounds and inverse-square-law tests~\cite{Lee:2020zjt} provide comparable constraints, both implying
$b R_0 \lesssim \mathcal{O}(10)\,\mu\textrm{m}$. These limits do not yet exclude the parameter region where $b R_0$ is of order $1\,\mu\textrm{m}$ as suggested by the Dark Dimension scenario~\cite{Montero:2022prj,Vafa:2024fpx}, though the allowed window is rather narrow and future improvements in short-distance gravity tests and astrophysical observations may decisively probe this regime.

\section{Casimir-induced vacuum energy
 in Dark Dimension}
\label{sec:casimir-stabilization}

We now turn to a concrete realization of the Dark Dimension scenario~\cite{Montero:2022prj}. In this framework, the inverse compactification radius is taken to be of order the sub-eV scale, which is close to the observed dark-energy scale,
\begin{equation}
    \rho_{\rm dark}^{1/4} \sim \mathcal{O}(10^{-3}\,\text{eV}) \,.
\end{equation}
This proximity suggests that the vacuum energy associated with the compact extra dimension, in particular, the \ac{kk} Casimir energy of light bulk fields, can naturally play an essential role in setting both the size of the extra dimension and the effective dark energy in four dimensions.

We consider a $5$-dimensional setup consisting of the bulk gravitational sector with a single compact extra dimension, a brane-localized \ac{sm}, and light bulk fermions that contain the right-handed neutrinos. In particular, we assume that the only fields beyond the gravitational sector that propagate in the bulk are three generations of right-handed neutrinos. This choice is motivated by the expectation that the compactification scale is of the same order as the neutrino mass scale.
The bulk sector contains only the massless $5$-dimensional graviton, which has five physical polarization states, so the number of bosonic degrees of freedom is $n_B = 5$.\footnote{We assume that supersymmetry is broken at a sufficiently large scale so that the graviphoton and gravitino do not contribute as light bulk degrees of freedom.} A $5$-dimensional Dirac fermion carries four degrees of freedom, so the total number of bulk fermionic degrees of freedom is $n_F = \sum_{i=1}^{3} n_F^{i} = 3 \times 4 = 12$.

The total \ac{kk} Casimir energy density can be written as
\begin{align}
    \rho_{\text{C}}(b)&=-\frac{3n_B\zeta (5)}{128 \pi ^7 (bR_0)^5}+\sum_{i=1}^3\frac{2n_F^iM_{i}^{5}}{(2 \pi)^{\frac{5}{2}}}
    \sum_{m=1}^{\infty} \frac{K_{\frac{5}{2}}\left(2\pi bR_0 M_{i}m\right)}{\left(2\pi bR_0M_{i}m\right)^{\frac{5}{2}}}\,
    \nonumber\\
    &= -\frac{3n_B \zeta (5)}{128 \pi ^7 (bR_0)^5}+\sum_{i=1}^3 \frac{2n_F^iM_{i}^{5}}{(2 \pi)^{\frac{5}{2}}} \sqrt{\frac{\pi}{2}} \left( \alpha_i^{-3} \mathrm{Li}_3 (e^{-\alpha_i}) + 3 \alpha_i^{-4} \mathrm{Li}_4 (e^{-\alpha_i}) \right.\notag\\
    &\quad + \left.3 \alpha_i^{-5} \mathrm{Li}_5 (e^{-\alpha_i}) \right),
    \label{eq:Casimir-energy-D5}
\end{align}
with
\begin{equation}
  \alpha_i \equiv 2 \pi b R_0 M_i\,,
\end{equation}
where $M_i$ ($i=1,2,3$) are the bulk neutrino masses, $\zeta(z)$ is the Riemann zeta function, $K_\nu(z)$ is the 
modified Bessel function of the second kind, and $\mathrm{Li}_s(z) \equiv \sum_{k=1}^{\infty} z^k/k^s$ is the polylogarithm. Appendix~\ref{app:KK-Casimir} gives the detailed derivation. 
The corresponding pressures in the three large spatial directions and in the extra dimension are
\begin{equation}
  p_{\text{C},a} (b) = - \rho_{\text{C}}(b), \quad p_{\text{C},b} (b) = - \frac{\partial}{\partial b}\bigl(\rho_{\text{C}} (b)\, b\bigr)\,,
\end{equation}
respectively. 
The first term in Eq.~\eqref{eq:Casimir-energy-D5} represents the contribution from the massless graviton, while the second term is the sum of the contributions from the three massive neutrino species. The massive contribution is exponentially suppressed for large $\alpha_i$ due to the asymptotic behavior of the polylogarithm, $\mathrm{Li}_s(z) \sim z$ as $|z| \to 0$.

In the small $b$ region~(i.e., $bR_0 M_i \ll 1$), the Casimir energy density can be expanded in powers of $bR_0M_i$, yielding
\begin{align}
\label{eq:Casimir-energy-D5-massless}
\rho_{\text{C}}(b) &\approx
-\frac{3n_B\zeta(5)}{128\pi^7 (bR_0)^5}
+ \sum_{i=1}^3\frac{3n_F^i}{128 \pi^7 } \left[\frac{\zeta(5)}{(bR_0)^5}
- \frac{(2\pi M_i)^2\zeta(3)}{6(bR_0)^3}
+ \cdots \right]\,.
\end{align}
Since $n_F > n_B$ in our setup, $\rho_{\text{C}}(b)$ diverges positively as $b \to 0^{+}$.

In the large $b$ region (i.e., $bR_0M_i \gg 1$), the massive contributions become exponentially suppressed, and the total Casimir energy density behaves as
\begin{equation}
    \label{eq:Casimir-energy-D5-massive}
    \rho_{\text{C}}(b) \simeq -\frac{3n_B \zeta (5)}{128 \pi ^7 (bR_0)^5} + \mathcal{O}\!\left(e^{-\alpha_{\mathrm{min}}}\right),
\end{equation}
implying that $\rho_{\text{C}}(b)$ necessarily approaches zero from the negative side at large $b$. Here, $\alpha_{\mathrm{min}}$ is defined as $\alpha_{\mathrm{min}} \equiv \min_i (\alpha_i)$.

By employing the expressions in both regimes~\eqref{eq:Casimir-energy-D5-massless} and~\eqref{eq:Casimir-energy-D5-massive}, we find that, in the absence of the $5$-dimensional cosmological constant $\Lambda_5$, the potential attains a negative minimum and does not provide a positive region, 
which is necessary for the realization of the small positive dark energy. 
This indicates that the minimal model considered in this section cannot explain the current accelerating expansion of the universe.
Since the Casimir energy for each particle species exhibits monotonic behavior, obtaining a positive contribution requires introducing a suitable number of massive bosonic and fermionic fields. To realize a positive dark energy, we therefore have to introduce either a tiny cosmological constant or extend the particle content of the minimal model.

In this work, we restrict our attention to the latter possibility, 
and assume that $\Lambda_5=0$. 
Introducing the cosmological constant would amount to postulating an additional finely tuned parameter, thereby obscuring the dynamical origin of the dark energy that we aim to explore here. By contrast, extending the particle content enables the vacuum energy to arise dynamically via the \ac{kk} Casimir effect, whose magnitude and sign are governed by physically well-defined properties such as particle masses, statistics, and boundary conditions.

\section{Dark Dimension quintessence and cosmological constraints}
\label{sec:dynamical-dark-energy}

In this section, we analyze the cosmological dynamics of the radion as a quintessence field and confront the Dark Dimension scenario with observational constraints. Our starting point is the $4$-dimensional effective theory obtained by dimensional reduction of the higher-dimensional setup. Since we are interested in the late-time homogeneous background evolution and in the time dependence of the radion $b(x)$, we restrict ourselves to the zero-modes of the bulk fields, integrate over the compact extra dimension, and neglect \ac{kk} excitations.

The Jordan-frame action~\eqref{eq:Jordan-frame} contains a non-minimal coupling between the radion $b(x)$ and the 4-dimensional Ricci scalar ${\cal R}(x)$. To bring the gravitational sector to canonical Einstein–Hilbert form, we perform a Weyl rescaling to the Einstein frame and introduce a canonically normalized radion field,
\begin{equation}
\label{eq:conformal-transformation}
g_{\mu\nu} = b^{-1}\,g^{\,(\text{E})}_{\mu\nu}, 
\qquad
\phi=\frac{M_{\rm Pl}\sqrt{3}}{4\sqrt{\pi}}\ln b\,,
\end{equation}
where $g^{(\text{E})}_{\mu\nu}$ denotes the Einstein-frame metric. After the rescaling, the effective 4-dimensional action
takes the following form
\begin{equation}
S_{\mathrm{eff}} 
= \int d^{4}x \sqrt{-g_{\text{E}}} 
\left[
\frac{M_{\mathrm{Pl}}^{2}}{16\pi} {\cal R}_{\text{E}}
- \frac{1}{2} (\partial_{\text{E}} \phi)^{2}
- V_{\mathrm{eff}}(\phi)
\right]
+ S_{\text{brane}}^{\text{SM}} + S_{\text{brane}}^{\text{DM}} \,,
\label{eq:Seff}
\end{equation}
where ${\cal R}_{\text{E}}$ is the Einstein-frame Ricci scalar, $(\partial_{\text{E}} \phi)^{2}\equiv g^{\mu\nu}_{\text{E}}\partial_{\mu}\phi\,\partial_{\nu}\phi$, and the brane-localized matter actions~$S_{\text{brane}}^{\text{SM}}$ and $S_{\text{brane}}^{\text{DM}}$ are rewritten in terms of the Einstein-frame metric and generally acquire radion-dependent couplings.
The effective potential is given by
\begin{align}
V_{\mathrm{eff}}(\phi)
=
{\cal V}\,\rho_{\text{C}}\big(b(\phi)\big)
\exp\!\left(
 -\frac{4\phi}{M_{\mathrm{Pl}}} \sqrt{\frac{\pi}{3}}
\right),
\label{eq:Veff_phi}
\end{align}
where $\rho_{\text{C}}(b)$ is the Casimir energy density.

In addition to the particle contents discussed in the previous section, 
we introduce two massive gauge bosons ($\Delta n_{\text{B}}^a = 4; a = 1, 2$) and two massless fermions ($\Delta n_{\text{F}} = 8$). Hence, the Einstein-frame radion potential can be written explicitly as
\begin{align}
V_{\mathrm{eff}}(\phi)
&= \frac{2\pi R_0}{b(\phi)}
\biggl[\rho_{\text{C}}(b, M_i) + \frac{3n_{\text{F}}\zeta (5)}{128 \pi ^7 (b(\phi)R_0)^5}
- \sum_{a=1}^2 \frac{2n_{\text{B}}^aM_{\text{B}, a}^{5}}{(2 \pi)^{\frac{5}{2}}}
\sum_{m=1}^{\infty}
\frac{
K_{\frac{5}{2}}\!\left(2\pi b(\phi)R_0 M_{\text{B}, a}m\right)
}{
\left(2\pi b(\phi)R_0M_{\text{B}, a}m\right)^{\frac{5}{2}}
}
\biggr]\,,
\label{eq:Veff_b-D5}
\end{align}

where $\rho_\text{C}(b, M_i)$ is the Casimir energy density 
in Eq.~\eqref{eq:Casimir-energy-D5},\footnote{Here we explicitly express the bulk neutrino masses~$M_i$ ($i=1,2,3$) as the second argument. } 
and $M_{\text{B},a}$ ($a=1,2$) are the masses of the gauge bosons. 
We choose the mass parameters as
\begin{align}
    (M_1R_0,\,M_2R_0,\,M_3R_0) &= (1,\, 3,\, 3) ,\nonumber\\ 
    (M_{\text{B}, 1} R_0,\, M_{\text{B},2} R_0) &= (0.725,\, 0.8). 
\end{align}
The massive gauge bosons introduced here may be regarded as effective degrees of freedom arising from a more complete \ac{uv} theory, for example via the Higgs or the St\"uckelberg mechanism. Since our focus is on the resulting low-energy Casimir contribution, the detailed \ac{uv} completion will not be specified in this work.

\begin{figure}
    \centering
    \includegraphics[width=0.8\linewidth]{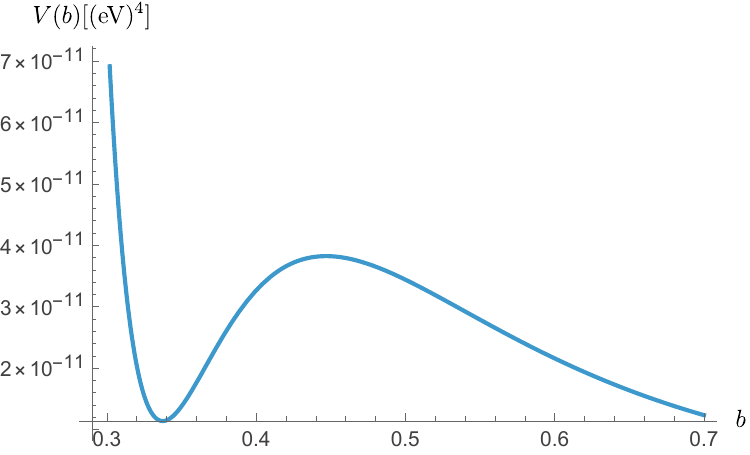}
    \caption{The effective potential determined by the Casimir energy.
The additional particle content in the 5-dimensional theory consists of two massive gauge bosons and two massless fermions.
In this figure, we take $R_0 = 83\, \textrm{eV}^{-1}=16.4\,\mu\textrm{m}$, which corresponds to choosing
$R_0 = \lambda \Lambda_{\rm c.c.}^{-1/4}$ with $\lambda \sim 10^{-1}$ \cite{Montero:2022prj}. The vertical axis represents $V(b) \,[\text{eV}^4]$.}
    \label{fig:effective-potential}
\end{figure}

Through the relation $b(\phi) \propto \exp\bigl(4\sqrt{\pi/3}\,\phi/M_{\rm Pl}\bigr)$, the radion potential $V_{\text{eff}}(\phi)$ acquires an overall exponential dependence on $\phi$, modulated by powers and the Bessel functions. Such potentials are typically steep and tend to drive a rather rapid evolution of the scalar field, making it non-trivial to realize slow-roll dynamics compatible with either inflation or late-time accelerated expansion. For example, the potential has an overall $b^{-4}$ dependence for small $b$, and the slow-roll condition is violated. In practice, by arranging for the effective potential to acquire a hilltop structure due to the Casimir contributions from various bulk fields, we can obtain a potential that meets the slow-roll requirements. 
A typical shape of $V_{\mathrm{eff}}(\phi)$ is illustrated in Fig.~\ref{fig:effective-potential}. To ensure $\phi>0$, we rescale $b$ in this figure. Specifically, we define the dimensionless variable $b \to \frac{b}{b_0}=\exp\bigl(4\sqrt{\pi/3}\,\phi/M_{\rm Pl}\bigr)$,  with $b_0=10^{-3}$.

For late-time cosmology, we consider non-relativistic matter (denoted by the subscript $\mathrm{m}$ below), namely, baryons and dark matter localized on separate 4-dimensional branes, i.e., on spacetime with three expanding spatial dimensions. In the higher-dimensional Jordan frame, this component therefore obeys the standard scaling $\rho_\textrm{m}^{(\rm{J})} \propto a_{(\textrm{J})}^{-3}$, where $a_{(\textrm{J})}(t)$ is the Jordan-frame scale factor. After the dimensional
reduction and the Weyl rescaling~\eqref{eq:conformal-transformation} to the Einstein frame, the matter fields become non-minimally coupled to the radion through the conformal factor. As a consequence, the Einstein-frame matter energy density becomes $\phi$-dependent explicitly
(see, e.g., Refs.~\cite{DeFelice:2010aj,Deruelle:2010ht,Chiba:2013mha}), 
which is parametrized as 
\begin{equation}
\rho_{\rm m}(a,\phi) = \sum_{i = \mathrm{b}, \mathrm{DM}} \rho_{{\rm i},0}\left(\frac{a_0}{a}\right)^{3}\exp\!\left(-c_i\,\phi\right),
\label{eq:rho_m_Einstein}
\end{equation}
where $a$ now denotes the Einstein-frame scale factor, and the subscripts b, DM, and $0$ denote baryon, dark matter, and the corresponding present-day value, respectively. The constant $c_i$ encodes the effective coupling in the Einstein frame and depends on how the brane sector couples to the higher-dimensional metric and to the radion.

If both baryons and dark matter are minimally coupled to the same induced metric, $c$ is not arbitrary but is fixed by the conformal factor associated
with the Weyl transformation. In the simplest realization,
this yields $c_\text{b} = c_\text{DM} =\frac{2}{M_{\mathrm{Pl}}} \sqrt{{\pi}/{3}}$.  Such a
universal coupling is tightly constrained by the fifth force and can also be constrained by astrophysical considerations, e.g., from the
dynamics of dwarf galaxies in the Milky Way environment~\cite{Kesden:2006zb}, suggesting that $c_\text{DM}$ is less than $\mathcal{O}(10^{-1})$. For \ac{sm} particles ($c_\text{b}$), however, analogous constraints are much tighter.  In the following, we neglect $c_\text{b}$ and vary $c \equiv c_\text{DM}$.

However, when the 5-dimensional action is derived from a higher-dimensional superstring, the coupling constant $c$ generally depends on the details of the compactification and brane construction, and may therefore take a range of 
values.
\footnote{
The simplest way to eliminate the interaction between matter and the radion arising from Weyl scaling is to assume that the total volume of all extra dimensions remains constant~\cite{Bedroya:2025fwh}. In that case, while the heavy higher-dimensional modes are expected to decouple at low energies, their effects can still survive through induced corrections to the radion potential.
}
Specifically, the brane metric may be expressed as $g^{\rm ind}_{\mu\nu} = C(\phi) g_{\mu\nu}$.\footnote{Since we consider the flat geometry for the extra dimension, $g_{\mu\nu}$ and $\phi$ are independent of $y$. }
Thus, the constant $c$ is controlled by $C(\phi)$. 
One may also consider a more general coupling of the dark matter sector to the radion through $g^{\rm ind}_{\mu\nu} = C(\phi) g_{\mu\nu} + D(\phi)\partial_\mu\phi\partial_\nu\phi$, which includes disformal terms~\cite{Koivisto:2013fta}. Such generalized couplings can modify the effective strength of the radion-dark-matter interaction, and the cosmological constraints on the conformal coupling may be relaxed~\cite{vandeBruck:2015ida}. In the present work, however, we restrict ourselves to a simpler model with the coupling parameter $c$, rather than analyzing the full induced metric with general $C(\phi)$ and $D(\phi)$. Based on these considerations, we examine a scenario in which the \ac{sm} and the dark matter are localised on separate branes and the dark matter sector possesses a phenomenologically valid coupling with the radion, $c$. In the context of the Dark Dimension, there is an argument that when $c$ is of the order of $\mathcal{O}(10^{-2})$, the interacting dark energy model is realized and explains observations such as those from DESI~\cite{Bedroya:2025fwh}. Hereafter, we will explore this possibility.

The radion field $\phi$ described by the effective 4-dimensional action~\eqref{eq:Seff} plays the role of a dynamical dark energy component. Specializing to a spatially homogeneous background, the Einstein equations reduce to the Friedmann equation
\begin{align}
\label{eq:Friedmann-t}
H^2
= \frac{8\pi}{3 M_\text{Pl}^2}
\left(
\frac{1}{2} \dot{\phi}^2
+ V_\text{eff}(\phi)
+ \rho_\text{m}(a,\phi)
\right),
\end{align}
where $H \equiv \dot{a}/a$ is the Einstein-frame Hubble parameter. The homogeneous radion satisfies
\begin{equation}
\label{eq:phiEOM-t}
\ddot{\phi}
+ 3 H \dot{\phi}
+ \frac{d V_\text{eff}(\phi)}{d\phi}
+ \frac{\partial \rho_\text{m}(a,\phi)}{\partial \phi}
= 0\,,
\end{equation}
where the last term reflects the explicit $\phi$-dependence of the dark matter sector and implies a non-trivial energy exchange between $\phi$ and dark matter.

The energy density and the pressure of the radion are defined as
\begin{equation}
\rho_\phi = \frac{1}{2}\dot\phi^2 + V_\text{eff}(\phi),
\qquad
p_\phi = \frac{1}{2}\dot\phi^2 - V_\text{eff}(\phi),
\end{equation}
so that the radion equation-of-state parameter is
\begin{equation}
w_\phi \equiv \frac{p_\phi}{\rho_\phi}
= \frac{\frac{1}{2}\dot\phi^2 - V_\text{eff}(\phi)}
       {\frac{1}{2}\dot\phi^2 + V_\text{eff}(\phi)}\,.
\end{equation}

Since there are interactions between $\phi$ and the cold dark matter, the cold dark matter density does not scale as $a^{-3}$, unlike in the case of the standard cold dark matter. To express the Friedmann equation in a form comparable to the standard non-interacting case, it is useful to define an effective dark-energy density and the equation-of-state parameter~\cite{Das:2005yj} (see also Refs.~\cite{Huey:2004qv, Bedroya:2025fwh}), 
\begin{align}
    \rho_\mathrm{eff} \equiv & \rho_\phi + \frac{\rho_{\mathrm{m},0}}{a^3} \left( e^{-c (\phi - \phi_0)} - 1\right),  \\
    w_\mathrm{eff} \equiv & \frac{w_\phi}{1 + \frac{\rho_{\mathrm{m},0}}{a^3 \rho_\phi} \left( e^{-c\left(\phi-\phi_0\right)} - 1\right)}.
\end{align}
These represent how the dark energy looks like when one neglects the $\phi$-dependence of $\rho_\text{m}$. 
Notice that $w_\text{eff}$ can take a phantom value $< -1$ even when $w_\phi \geq -1$.

For the numerical analysis and a comparison with cosmological observations, we solve the background evolution in terms of the e-folding number $N \equiv \ln(a/a_0)$, where $N=0$ today and the redshift is given by $1+z=e^{-N}$. The explicit form of the Friedmann and the radion equations in terms of $N$ is given in Appendix~\ref{app:efolding}. Once a consistent background solution is obtained, the resulting equation-of-state parameter $w_\phi(z)$ can be compared with observational reconstructions. 

 \begin{figure}[tp]%
\centering%
\includegraphics[width=\linewidth]{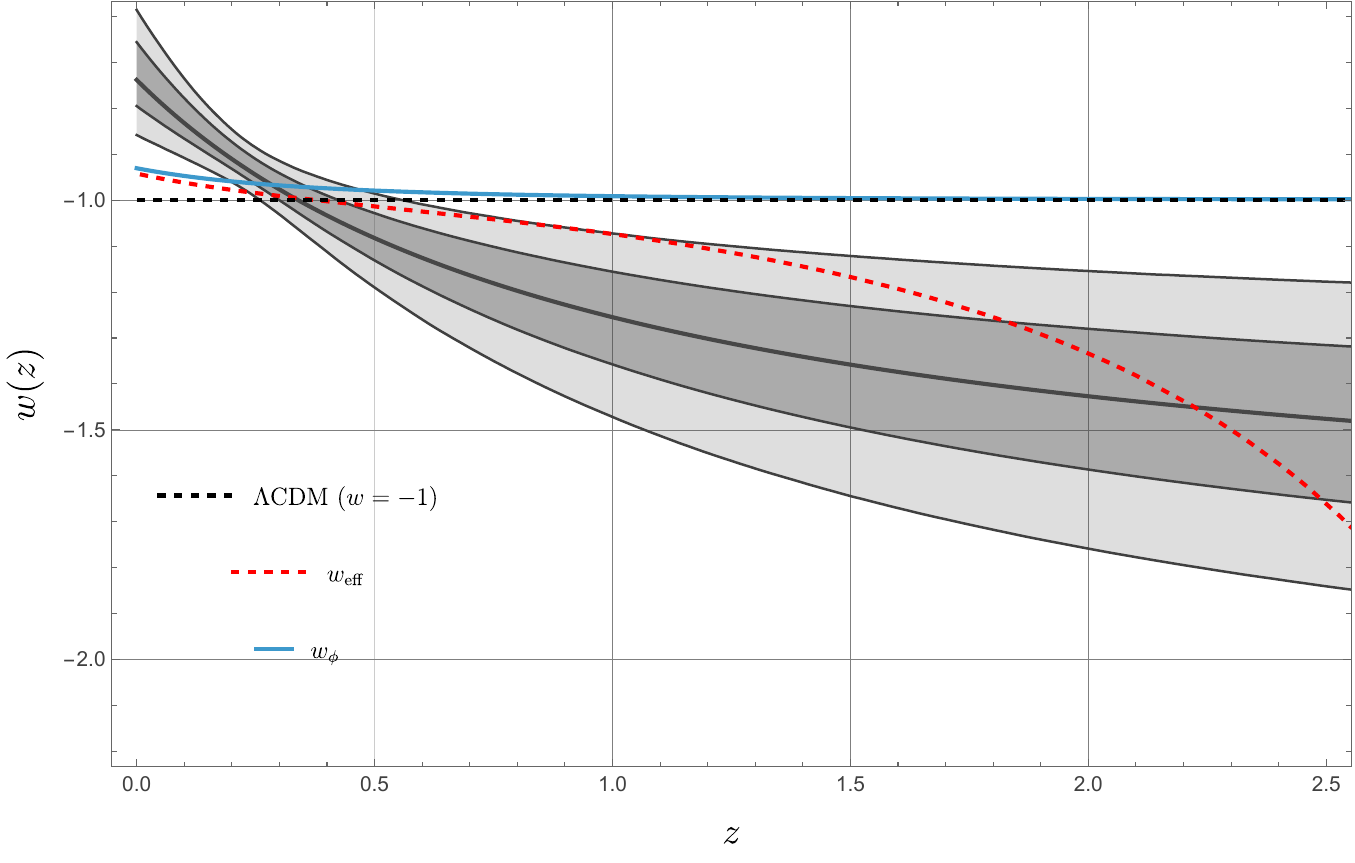}%
\caption{Equation-of-state parameter $w(z)$ for the Dark-Dimension model. The red dashed line and blue solid line represent $w_{\rm eff}(z)$ and $w_{\phi}(z)$ for $c=0.005$. The black solid line and gray area represent the constraints of $w(z)$ for CPL model using DESI+CMB+DESY5~\cite{DESI:2025zgx}.}
\label{fig:weff}
\end{figure}
 
 An example of the radion equation-of-state parameter for the dark-dimension model is shown in Fig.~\ref{fig:weff}. The gray areas in Fig.~\ref{fig:weff} are constraints obtained using DESI+CMB+DESY5 observational data on the Chevallier-Polarski-Linder (CPL) model \cite{Chevallier:2000qy,Linder:2002et} in Ref.~\cite{DESI:2025zgx}. The equation of state of the CPL model is defined as
\begin{align}
    \label{eq:CPL}
    w(z)=w_0+w_a\frac{z}{1+z},
\end{align}
where $w_0$ is the present value of the equation-of-state parameter and $w_a$ characterizes its time evolution. Recent DESI observations suggest a possible preference for dynamical dark energy over the $\Lambda$CDM model. The CPL parametrization is widely used to describe such an evolving dark energy scenario and provides a convenient phenomenological fit to observational data. In particular, the region of the $(w_0,w_a)$ parameter space favored by these observations typically corresponds to an evolution of the dark energy equation of state that crosses the phantom divide $w=-1$. Such a phantom crossing behavior is difficult to realize in many simple physical models. Therefore, constructing physically motivated models that can reproduce a CPL-like dynamical dark energy evolution, including the possibility of phantom crossing suggested by the data, is an important task. Fig.~\ref{fig:weff} shows that, except near $z\sim0$, the Dark Dimension model reproduces the DESI-preferred dynamical dark energy equation of state and qualitatively agrees with the best-fit CPL behavior. In our model, the dark energy equation-of-state parameter crosses the phantom divide at $z\sim0.4$, consistent with the redshift range suggested by the DESI preference for dynamical dark energy. 

\begin{table}
    \centering
    \caption{A summary of the observational data from DESI DR2~\cite{DESI:2025zgx}  with $r_d=147.1$~Mpc.  }
    \label{tab:DESIDR2_data}
    \begin{tabular}{|c|c|c|} \hline
         $z_{\rm eff}$ & $D_M/r_d$ & $D_H/r_d$  \\ \hline
         $0.51$ & $13.59\pm0.17$ & $21.86\pm0.43$ \\ \hline
         $0.71$ & $17.35\pm0.18$ & $19.46\pm0.33$  \\ \hline
         $0.93$ & $21.58\pm0.15$ & $17.64\pm0.19$  \\ \hline
         $1.32$ & $27.60\pm0.32$ & $14.18\pm0.22$  \\ \hline
         $1.48$ & $30.51\pm0.76$ & $12.82\pm0.52$  \\ \hline
         $2.33$ & $38.99\pm0.94$ & $8.63\pm0.10$  \\ \hline
    \end{tabular}
\end{table}

\begin{figure}
    \centering
    \includegraphics[width=1.0\linewidth]{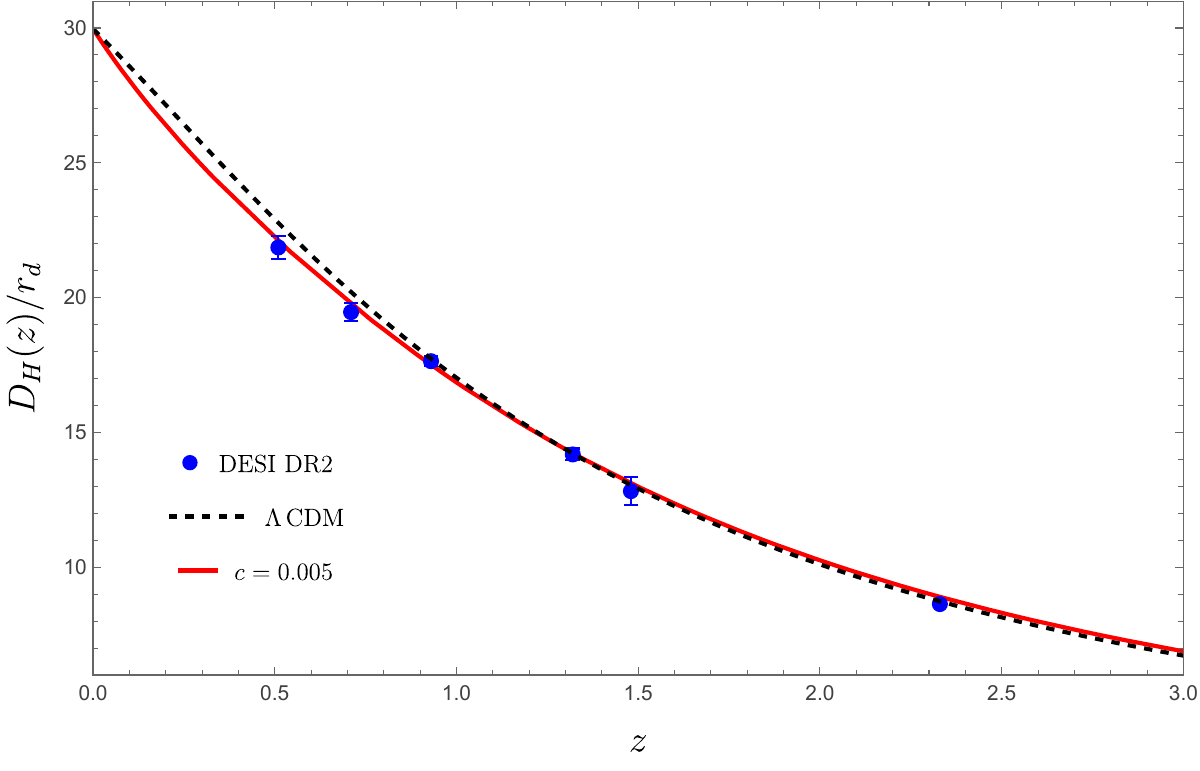}
    \caption{This plot represents the BAO observable $D_H/r_d$. The black dashed and the red solid lines represent the $\Lambda$CDM ($\Omega_{\rm m,0}=0.3,~\Omega_{\Lambda}=0.7$) and the Dark-Dimension ($c=0.005$) 
    model, respectively. Blue points and error bars are observational data from DESI DR2 summarized in Tab.~\ref{tab:DESIDR2_data}. In this analysis, we used $r_d=147.1~{\rm Mpc}$. 
    }
    \label{fig:DH_rd}
\end{figure}

\begin{figure}
    \centering
    \includegraphics[width=1.0\linewidth]{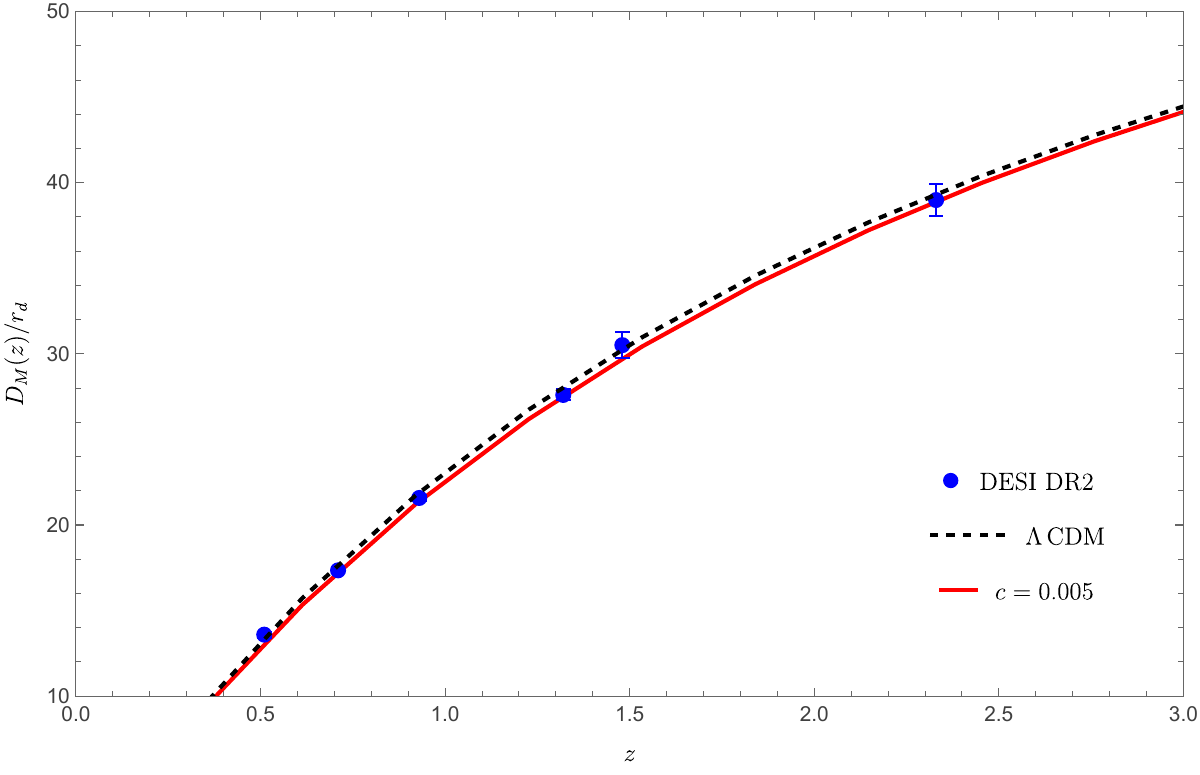}
    \caption{This plot represents the BAO observable $D_M/r_d$. The black dashed and the red solid line represent the $\Lambda$CDM ($\Omega_{\rm m,0}=0.3,~\Omega_{\Lambda}=0.7$) and the Dark-Dimension ($c=0.005$) 
    model, respectively. Blue points and error bars are observational data from DESI DR2 summarized in Tab.~\ref{tab:DESIDR2_data}. In this analysis, we used $r_d=147.1~{\rm Mpc}$.}
    \label{fig:DM_rd}
\end{figure}

In the flat universe (the curvature density fraction $\Omega_K=0$), the Hubble distance $D_H(z)$, the comoving distance $D_C(z)$, and the (comoving) angular diameter distance $D_M(z)$ are given by
\begin{align}
\label{eq:DH_DC_DM}
D_H(z)=\frac{1}{H(z)}, \qquad
D_C(z)=\int_{0}^{z}\frac{dz'}{H(z')}, \qquad
D_M(z)=D_C(z).
\end{align}
We compare these with the DESI DR2 observational data shown in Table~\ref{tab:DESIDR2_data}. Figs.~\ref{fig:DH_rd} and \ref{fig:DM_rd} show the calculated $D_H(z)/r_d$ and $D_M(z)/r_d$ for the Dark Dimension model with the solid red lines. In this analysis, we fixed $r_d=147.1~{\rm Mpc}$ as the demonstration. To quantify the agreement with the DESI DR2 BAO measurements, we compute a simple diagonal $\chi^2$ defined as
\begin{align}
      \chi^2 = \sum_i \frac{(X_{\rm model}(z_i)-X_{\rm data}(z_i))^2}{\sigma_i^2},
\end{align}
where $X$ denotes either $D_H/r_d$ or $D_M/r_d$, and $\sigma_i$ is the quoted observational uncertainty. For the parameter set shown in Figs.~\ref{fig:DH_rd} and \ref{fig:DM_rd}, our model gives $\chi^2_{D_H}=9.28$ and $\chi^2_{D_M}=10.88$, while the corresponding values for $\Lambda$CDM are $\chi^2_{D_H}=10.69$ and $\chi^2_{D_M}=12.71$. The total $\chi^2$ is therefore $\chi^2_{\rm model}=19.97$ and $\chi^2_{\Lambda{\rm CDM}}=23.79$, corresponding to an improvement of $\Delta\chi^2 \simeq -4.0$. This example indicates that our model possesses parameter regions that can reproduce the DESI BAO measurements better than $\Lambda$CDM, at least within the diagonal $\chi^2$ approximation.

\section{Conclusions}
\label{sec:conclusion}
In this work, we investigated a concrete realization of the Dark Dimension scenario in which a single large extra dimension is in the sub-millimeter range and the associated Casimir energy contributes to the observed dark energy. We considered a 5-dimensional setup in which the \ac{sm} fields are localized on a 4-dimensional brane, while gravity and a small number of light fields propagate in the bulk. Starting from the higher-dimensional theory, we derived the 4-dimensional effective action for the radion and showed that it can play the role of a quintessence dark energy.

We computed the \ac{kk} Casimir energy and analyzed the resulting radion potential. In the minimal model containing only gravity and three bulk neutrinos, the radion potential develops a negative minimum, and this setup cannot account for a positive late-time vacuum energy. We therefore considered extended particle content in the bulk and showed that the Casimir-induced potential can instead yield a positive region relevant for dark-energy phenomenology. In representative models, the corresponding compactification scale is of order $bR_0 \sim 1\,\mu\mathrm{m}$, consistent with the scale suggested in the Dark Dimension scenario, and implies a fundamental scale $M_5 \sim 10^9\,\mathrm{GeV}$.

After the dimensional reduction and the Weyl transformation to the Einstein frame, the radion is described by a canonically normalized scalar field with a Casimir-induced potential and an effective coupling to matter. We studied the resulting homogeneous cosmological evolution and obtained the redshift dependence of the equation-of-state parameter for representative parameter choices. As shown in Fig.~\ref{fig:weff}, our results show that the radion can realize quintessence behavior and can be compatible with current cosmological constraints, including recent DESI data. Furthermore, in Figs.~\ref{fig:DH_rd} and \ref{fig:DM_rd}, we compared the BAO observables $D_H/r_d$ and $D_M/r_d$ predicted by our model with those of the $\Lambda$CDM model. We found that the $\chi^2$ value for our model is smaller than that for $\Lambda$CDM. This indicates that our model possesses parameter regions that can reproduce the DESI BAO measurements better than $\Lambda$CDM, at least within the diagonal $\chi^2$ approximation. Our findings suggest that the Dark Dimension scenario provides a \ac{uv}-motivated framework for understanding the nature of dark energy through the quantum effects of higher-dimensional gravity.

\section*{Acknowledgments}
HM is supported by JSPS Grant-in-Aid for Scientific Research Number
23K13100. YM is supported by JST SPRING, Grant Number JPMJSP2110. TT is supported by the 34th Academic research grant FY 2024 (Natural Science) No.~9284 from DAIKO FOUNDATION. FO~is supported by individual research funding from Nihon University.

\appendix

\section{Kaluza-Klein Casimir energy on an $n$-torus}
\label{app:KK-Casimir}

In this appendix, we outline the derivation of the one-loop \ac{kk} Casimir energy for a bulk field in a $D$-dimensional spacetime compactified on an $n$-torus, $T^n$, closely following Refs.~\cite{Birrell:1982ix,Arkani-Hamed:2007ryu}. We adopt a semiclassical approach in which the spacetime metric is treated classically while quantum effects of matter fields are incorporated via the expectation value of the stress-energy tensor. This is reliable at low energies where quantum gravitational corrections are subdominant.
The $D$-dimensional semiclassical Einstein equations are
\begin{equation}
G_{\mu\nu}^{(D)} + \Lambda_{(D)} g_{\mu\nu}^{(D)} = 8\pi G_{(D)} \langle T_{\mu\nu}^{(D)} \rangle\,,
\label{eq:semiclassical-Einstein}
\end{equation}
where $\langle T_{\mu\nu}^{(D)} \rangle$ is the renormalised expectation value of the energy-momentum tensor for quantum fields. In this work we compute $\langle T_{\mu\nu}^{(D)} \rangle$ in the flat-spacetime limit, so higher-derivative curvature terms do not appear explicitly. In more general curved backgrounds, such terms would have to be treated within the framework of the effective field theory~\cite{Buchbinder:1992rb,Stelle:1976gc,Horowitz:1978fq,Horowitz:1980fj,Hartle:1981zt,RandjbarDaemi:1981wd,Jordan:1987wd,Suen:1988uf,Suen:1989bg,Anderson:2002fk,Matsui:2019tlf,Simon:1991bm,Parker:1993dk}.

In this appendix, we consider general $D$ for completeness. The discussion in the main text corresponds to $D=5$ and $n=1$. 

\subsection{Casimir energy from Green's functions}

For simplicity, we consider a real scalar field $\Phi$ in a $D$-dimensional spacetime with metric $g^{(D)}_{\mu\nu}$,
\begin{equation}
\mathcal{L}^{(D)}_m= -\frac{1}{2}g^{\mu\nu}_{(D)}\nabla_\mu\Phi\nabla_\nu\Phi-\frac{1}{2}M^2\Phi^2, 
\end{equation}
where $\mu,\nu=0,1,\dots,3+n$, $\nabla_\mu$ is the ($D$-dimensional) covariant derivative, and $M$ is the bulk mass parameter. The expectation value of the quantum energy-momentum tensor can be written as
\begin{align}\label{eq:quantum-energy-momentum-tensor}
\Braket{T_{\mu\nu}^{(D)}} & =\left\langle \mathcal{L}^{(D)}_m g^{(D)}_{\mu \nu}-2 \frac{\delta \mathcal{L}^{(D)}_m}
{\delta g^{(D)}_{\mu \nu}}\right\rangle \notag \\
& =\lim _{x^{\prime} \rightarrow x}\left(\frac{1}{2}\left(\nabla_\mu \nabla_\nu^{\prime}+\nabla_\nu \nabla_\mu^{\prime}\right)-\frac{1}{2} g^{(D)}_{\mu \nu}\left(\nabla^\alpha \nabla_\alpha^{\prime}+M^2\right)\right) G\left(x-x^{\prime}\right)\,,
\end{align}
where $G\left(x-x^{\prime}\right)=\left\langle\hat{\Phi}\left(x\right)\hat{\Phi}\left(x^{\prime}\right)\right\rangle$ is the free-field propagator, which depends only on the invariant distance between the two points, $\nabla^\mu = g^{(D)\mu \nu}\nabla_{\nu}$, and $\nabla'_\mu$ ($\nabla_\mu$) is the covariant derivative with respect to $x'$ ($x$).

We compactify $n$ spatial dimensions on an isotropic torus $T^n$, which has a common coordinate radius~$R_0$ and physical radius $b R_0$. Denoting the coordinates of the compact dimensions by $y_i$ ($i=1,\dots,n$) with periodicity $0\leq y_i<2\pi R_0$, the compactification can be implemented via the method of images, which expresses the propagator on $T^n$ as a sum over images of the propagator in uncompactified flat space~\cite{Yu:1999pq,Yu:2000yf}:
\begin{equation}
G_\text{C}\left(x-x^{\prime}\right)=\sum_{\{m_i\}\neq 0}G\left(x-x^{\prime}+2 \pi bR_0 m_1|_{y_1}+\cdots +2 \pi bR_0 m_n|_{y_n}  \right) , 
\label{def:G_C}
\end{equation}
where the integers $m_i$ label the images along each compact direction, and we omitted the contribution $m_1 = m_2 = \cdots = m_n = 0$ and denoted them by $\{m_i \} \neq 0$.

Substituting $G_\text{C}$ into Eq.~\eqref{eq:quantum-energy-momentum-tensor} and subtracting the uncompactified contribution yields the renormalised Casimir energy-momentum tensor.

\subsection{Flat-spacetime approximation}

In flat spacetime, the propagator for a massive scalar field is
\begin{equation}
G\left(x-x^{\prime}\right)=\int \frac{d^D k}{(2 \pi)^D} \frac{e^{i k\left(x-x^{\prime}\right)}}{k^2+M^2}\,.
\end{equation}
Acting with $(\partial^\alpha \partial_\alpha^{\prime}+M^2)$ on $G$ gives
\begin{align}
\frac{g_{\mu \nu}^{(D)}}{2} \left(\partial^\alpha \partial_\alpha^{\prime}+M^2\right) G\left(x-x^{\prime}\right) 
&= \frac{g^{(D)}_{\mu\nu}}{2}\int\frac{d^Dk}{(2\pi)^D}e^{ik(x-x')}=\frac{g^{(D)}_{\mu\nu}}{2}\delta^{(D)} (x-x')\,.
\end{align}
Thus, from Eq.~\eqref{def:G_C}, we have
\begin{align}
     &\quad\frac{g_{\mu\nu}^{(D)}}{2}\left(\partial^\alpha\partial_\alpha^\prime+M^2\right)G_\text{C}(x-x')
     \nonumber\\
     &= \frac{g_{\mu\nu}^{(D)}}{2}\sum_{\{m_i\}\neq 0}\delta^{(D)}\left(x-x'+2\pi bR_0m_1|_{y_1}+\cdots+2\pi bR_0m_n|_{y_n}\right)\,.
\end{align}
Note that these delta functions vanish in the coincidence limit $x\to x'$ 
when at least one $m_i\neq 0$ ($i=1,\cdots,n$). 
As a result, only the first term in Eq.~\eqref{eq:quantum-energy-momentum-tensor} contributes to the Casimir energy-momentum tensor, which can be written as
\begin{align}
\Braket{T_{\mu\nu}^{(D)}}_\text{C} &=\lim _{x^{\prime} \rightarrow x}\left(\frac{1}{2}\left(\partial_\mu \partial_\nu^{\prime}+\partial_\nu \partial_\mu^{\prime}\right)\right)\sum_{\{m_i\}\neq 0}G_\text{C}\left(x-x^{\prime}+2 \pi bR_0 m_1|_{y_1}\cdots +2 \pi bR_0 m_n|_{y_n}  \right) \notag \\ 
&=- \sum_{\{m_i\}\neq 0}\lim _{x^{\prime} \rightarrow x}\partial_\mu \partial_\nu 
G_\text{C}\left(x-x^{\prime}+2 \pi bR_0 m_1|_{y_1}\cdots +2 \pi bR_0 m_n|_{y_n}  \right) \,,
\end{align}
where we used the relation $\partial_\mu \partial_\nu^{\prime}=-\partial_\mu \partial_\nu$
for $G\left(x-x^{\prime}\right)$.

For a massless scalar field in $D$ dimensions, the Green's function is
\begin{align}
G\left(r^2\right)=\frac{\Gamma\left(\frac{D}{2}-1\right)}
{4 \pi^{\frac{D}{2}}} \frac{1}{r^{D-2}}\,,
\end{align}
where $r^2$ is the squared distance between the two points. After some algebra, we find that the Casimir energy density and pressures take the form
\begin{align}
\begin{split}
\rho_{\text{C}} &= \frac{\Gamma\left(\frac{D}{2}-1\right)}{2\pi^{\frac{D}{2}}}
\left(\frac{2-D}{2}\right)
\sum_{\{m_i\}\neq 0}
\frac{1}{\left(2\pi bR_0\right)^D\left(m_1^2+\cdots +m_n^2\right)^{\frac{D}{2}}}\,, \\
p_{\text{C},a} &= -\rho_{\text{C}}\,,  \\ 
p_{\text{C},b} &= -\frac{\Gamma\left(\frac{D}{2}-1\right)}{2\pi^{\frac{D}{2}}}
\left(\frac{2-D}{2}\right)
\sum_{\{m_i\}\neq 0}
\frac{1}{\left(2\pi bR_0\right)^D\left(m_1^2+\cdots +m_n^2\right)^{\frac{D}{2}}}\\
& \quad + \frac{\Gamma\left(\frac{D}{2}-1\right)}{2\pi^{\frac{D}{2}}}
\left(\frac{D(2-D)}{2}\right)
\sum_{\{m_i\}\neq 0}
\frac{m_n^2}{\left(2\pi bR_0\right)^D\left(m_1^2+\cdots +m_n^2\right)^{\frac{D+2}{2}}}\,,    
\end{split}
\end{align}
where $p_{\text{C},a}$ is the pressure in the three large spatial dimensions and $p_{\text{C},b}$ is the pressure along one of the compact directions (by symmetry, all compact directions are equivalent).
For $D=5$ and $n=1$ (one compact dimension), the sums become the Riemann zeta functions, and we obtain
\begin{equation}
\rho_{C}
=-\frac{3 \zeta (5)}{128 \pi ^7 (bR_0)^5}\,, \qquad 
p_{C,b}=-\frac{3 \zeta (5)}{32 \pi ^7 (bR_0)^5}\,,
\end{equation}
where $\zeta(5)=1.036927\cdots$. These are precisely the massless graviton contributions used in Sec.~\ref{sec:casimir-stabilization}.

For a massive scalar field, the flat-space propagator can be written in terms of the modified Bessel function of the second kind,
\begin{equation}
G\left(r^2\right)=\frac{M^{D-2}}{(2 \pi)^{\frac{D}{2}}} 
\frac{K_{\frac{D}{2}-1}(M r)}{(M r)^{\frac{D}{2}-1}}\,.
\end{equation}
Using standard differential identities for $K_\nu(z)$, we find that the Casimir energy density and the pressures become
\begin{align}
\begin{split}
\rho_{\text{C}} &= -\frac{M^{D}}{(2 \pi)^{\frac{D}{2}}}
\sum_{\{m_i\}\neq 0}
\frac{K_{\frac{D}{2}}\left(2\pi bR_0 M\left(m_1^2+\cdots +m_n^2\right)^{\frac{1}{2}}\right)}{\left(2\pi bR_0 M\right)^{\frac{D}{2}}\left(m_1^2+\cdots +m_n^2\right)^{\frac{D}{4}}}\,, \\
p_{\text{C},a} &= -\rho_{\text{C}}\,,  \\ 
p_{\text{C},b} &= \frac{M^{D}}{(2 \pi)^{\frac{D}{2}}}
\sum_{\{m_i\}\neq 0}
\frac{K_{\frac{D}{2}}\left(2\pi bR_0 M\left(m_1^2+\cdots +m_n^2\right)^{\frac{1}{2}}\right)}{\left(2\pi bR_0 M\right)^{\frac{D}{2}}\left(m_1^2+\cdots +m_n^2\right)^{\frac{D}{4}}}\\
& \quad -\frac{M^{D+2}}{(2 \pi)^{\frac{D}{2}}}
\sum_{\{m_i\}\neq 0}
\frac{\left(2\pi bR_0M m_n\right)^2K_{\frac{D}{2}+1}\left(2\pi bR_0 M\left(m_1^2+\cdots +m_n^2\right)^{\frac{1}{2}}\right)}{\left(2\pi bR_0 M\right)^{\frac{D}{2}+1}\left(m_1^2+\cdots +m_n^2\right)^{\frac{D}{4}+\frac{1}{2}}}\,.    
\end{split}
\end{align}
For $D=5$ and $n=1$, this is simplified to
\begin{align}
\begin{split}
\rho_{\text{C}} &= -\frac{2M^{5}}{(2 \pi)^{\frac{5}{2}}}
\sum_{m=1}^{\infty}
\frac{K_{\frac{5}{2}}\left(2\pi bR_0 Mm\right)}{\left(2\pi bR_0 Mm\right)^{\frac{5}{2}}}\,, \\ 
p_{\text{C},b} &= \frac{2M^{5}}{(2 \pi)^{\frac{5}{2}}}
\sum_{m=1}^{\infty}
\frac{K_{\frac{5}{2}}\left(2\pi bR_0 Mm\right)}{\left(2\pi bR_0 Mm\right)^{\frac{5}{2}}}
-\frac{2M^{5}}{(2 \pi)^{\frac{5}{2}}}
\sum_{m=1}^{\infty}
\frac{\left(2\pi bR_0 Mm\right)^2K_{\frac{7}{2}}\left(2\pi bR_0 Mm\right)}{\left(2\pi bR_0 Mm\right)^{\frac{7}{2}}}\,.  
\end{split}
\end{align}
This expression for $\rho_{\text{C}}$ is the massive part of Eq.~\eqref{eq:Casimir-energy-D5} used in Sec.~\ref{sec:casimir-stabilization}.

\subsection{Example: 6-dimensional case with two extra dimensions}

For completeness, we also show the expressions in the case of $D=6$ and  $n=2$. In the minimal setup, the bulk field content consists of the 6-dimensional graviton and three generations of bulk fermions. 

In this case, the Casimir energy density for the compactified extra dimensions takes the form
\begin{align}
\begin{split}\label{eq:Casimir-energy-D6-n2}
\rho_{C}(b) &=-\frac{n_B}{16 \pi ^9 (bR_0)^6}\sum_{m_1=1}^{\infty}\sum_{m_2=1}^{\infty}
\frac{1}{\left(m_1^2+m_2^2\right)^{3}}\\
&\quad +\sum_{i=1}^3\frac{n_F^iM^{6}_i}{(2 \pi)^{3}}
\sum_{m_1=1}^{\infty}\sum_{m_2=1}^{\infty}
\frac{K_{3}\left(2\pi bR_0 M_i\left(m_1^2+m_2^2\right)^{\frac{1}{2}}\right)}{\left(2\pi bR_0 M_i\right)^{3}\left(m_1^2+m_2^2\right)^{\frac{3}{2}}}\,,
\end{split}
\end{align}
with analogous expressions for the pressures along the large and compact directions. The sums over $(m_1,m_2)$ are convergent and can be evaluated numerically; for the massless part we find, for example,
\begin{equation}
\sum_{m_1=1}^{\infty}\sum_{m_2=1}^{\infty}
\frac{1}{\left(m_1^2+m_2^2\right)^{3}}\approx 0.147385\,.
\end{equation} 
The qualitative behavior of $\rho_\text{C}(b)$ is similar to the $5$-dimensional case: the Casimir energy is positive at small radius (dominated by the fermionic contribution) and negative at large radius (dominated by the bosonic graviton contribution). However, as discussed in Sec.~\ref{sec:casimir-stabilization}, the phenomenology of the two-dimensional Dark-Dimension scenario is strongly constrained by the collider and the astrophysical bounds, and we have therefore focused on the $5$-dimensional case in our paper.

\section{Background cosmological equations for numerical analysis}
\label{app:efolding}

For a numerical analysis and a direct comparison with the cosmological data, it is convenient to reformulate the dynamics in terms of the e-folding number $N \equiv \ln\frac{a}{a_0}$, 
so that $a/a_0 = e^{N}$ and the redshift is given by $1+z = a_0/a = e^{-N}$. Derivatives with respect to the cosmic time and $N$ are related by $\frac{d}{dt} = H \frac{d}{dN}$, and we denote $d/dN$ by a prime, e.g. $\phi'(N) \equiv d\phi/dN$. In terms of $N$, the Friedmann equation~\eqref{eq:Friedmann-t} becomes
\begin{align}
H^2(N)
= \frac{8\pi}{3 M_\text{Pl}^2}
\left[
\frac{1}{2} H^2(N)\,\phi'(N)^2
+ V_\text{eff}(\phi(N))
+ \rho_\text{m}(N,\phi(N))
\right], \label{eq:Friedmann-N0}
\end{align}
where $\rho_\text{m}(N,\phi) \equiv \rho_\text{m}\bigl(a(N),\phi\bigr)$ follows from Eq.~\eqref{eq:rho_m_Einstein}. Solving Eq.~\eqref{eq:Friedmann-N0} for $H^2(N)$ yields
\begin{align}
H^2(N)
= \frac{\displaystyle \frac{8\pi}{3 M_\text{Pl}^2}
\Bigl[V_\text{eff}(\phi(N))
+ \rho_\text{m}(N,\phi(N))\Bigr]}
{1-\displaystyle \frac{4\pi}{3 M_\text{Pl}^2}\,\phi'(N)^2}\,.
\label{eq:Friedmann-N}
\end{align}
The radion equation of motion~\eqref{eq:phiEOM-t} can be rewritten as
\begin{equation}
\phi''(N)
+ \left(3 + \frac{H'(N)}{H(N)}\right)\phi'(N)
+ \frac{1}{H(N)^2}\frac{d V_\text{eff}(\phi)}{d\phi}
+ \frac{1}{H(N)^2}\frac{\partial \rho_\text{m}(N,\phi)}{\partial \phi}
= 0\,.
\label{eq:phiEOM-N}
\end{equation}
In terms of $N$, the equation-of-state parameter becomes
\begin{equation}
w_\phi(N)
= \frac{\frac{1}{2} H^2(N)\,\phi'(N)^2 - V_\text{eff}(\phi(N))}
       {\frac{1}{2} H^2(N)\,\phi'(N)^2 + V_\text{eff}(\phi(N))}\,.
\label{eq:wphi-N}
\end{equation}
Using Eqs.~\eqref{eq:Friedmann-N}, \eqref{eq:phiEOM-N} and $1+z=e^{-N}$, we obtain  the expression of $w_\phi$ as a function of the redshift~$z$.
The present-day Hubble parameter and the matter fraction are 
\begin{equation}
H(N=0) = H_0 \simeq 68\,\mathrm{km/s/Mpc},
\qquad
\Omega_{\mathrm{m},0} \simeq 0.3,
\end{equation}
so that $\Omega_{\phi,0} \simeq 0.7$ for a spatially flat universe.

\bibliography{Refs}

@article{Riess:1998cb,
    author = "Riess, Adam G. and others",
    collaboration = "Supernova Search Team",
    title = "{Observational evidence from supernovae for an accelerating universe and a cosmological constant}",
    eprint = "astro-ph/9805201",
    archivePrefix = "arXiv",
    doi = "10.1086/300499",
    journal = "Astron. J.",
    volume = "116",
    pages = "1009--1038",
    year = "1998"
}

@article{Perlmutter:1998np,
    author = "Perlmutter, S. and others",
    collaboration = "Supernova Cosmology Project",
    title = "{Measurements of $\Omega$ and $\Lambda$ from 42 High Redshift Supernovae}",
    eprint = "astro-ph/9812133",
    archivePrefix = "arXiv",
    reportNumber = "LBNL-41801, LBL-41801",
    doi = "10.1086/307221",
    journal = "Astrophys. J.",
    volume = "517",
    pages = "565--586",
    year = "1999"
}

@article{Spergel:2003cb,
    author = "Spergel, D. N. and others",
    collaboration = "WMAP",
    title = "{First year Wilkinson Microwave Anisotropy Probe (WMAP) observations: Determination of cosmological parameters}",
    eprint = "astro-ph/0302209",
    archivePrefix = "arXiv",
    doi = "10.1086/377226",
    journal = "Astrophys. J. Suppl.",
    volume = "148",
    pages = "175--194",
    year = "2003"
}

@article{Komatsu:2010fb,
    author = "Komatsu, E. and others",
    collaboration = "WMAP",
    title = "{Seven-Year Wilkinson Microwave Anisotropy Probe (WMAP) Observations: Cosmological Interpretation}",
    eprint = "1001.4538",
    archivePrefix = "arXiv",
    primaryClass = "astro-ph.CO",
    doi = "10.1088/0067-0049/192/2/18",
    journal = "Astrophys. J. Suppl.",
    volume = "192",
    pages = "18",
    year = "2011"
}

@article{Weinberg:1988cp,
    author = "Weinberg, Steven",
    editor = "Hsu, Jong-Ping and Fine, D.",
    title = "{The Cosmological Constant Problem}",
    reportNumber = "UTTG-12-88",
    doi = "10.1103/RevModPhys.61.1",
    journal = "Rev. Mod. Phys.",
    volume = "61",
    pages = "1--23",
    year = "1989"
}

@article{Zlatev:1998tr,
    author = "Zlatev, Ivaylo and Wang, Li-Min and Steinhardt, Paul J.",
    title = "{Quintessence, cosmic coincidence, and the cosmological constant}",
    eprint = "astro-ph/9807002",
    archivePrefix = "arXiv",
    doi = "10.1103/PhysRevLett.82.896",
    journal = "Phys. Rev. Lett.",
    volume = "82",
    pages = "896--899",
    year = "1999"
}

@article{Steinhardt:1999nw,
    author = "Steinhardt, Paul J. and Wang, Li-Min and Zlatev, Ivaylo",
    title = "{Cosmological tracking solutions}",
    eprint = "astro-ph/9812313",
    archivePrefix = "arXiv",
    doi = "10.1103/PhysRevD.59.123504",
    journal = "Phys. Rev. D",
    volume = "59",
    pages = "123504",
    year = "1999"
}

@book{Green:1987sp,
    author = "Green, Michael B. and Schwarz, J. H. and Witten, Edward",
    title = "{SUPERSTRING THEORY. VOL. 1: INTRODUCTION}",
    isbn = "978-0-521-35752-4",
    series = "Cambridge Monographs on Mathematical Physics",
    month = "7",
    year = "1988"
}

@book{Polchinski:1998rq,
    author = "Polchinski, J.",
    title = "{String theory. Vol. 1: An introduction to the bosonic string}",
    doi = "10.1017/CBO9780511816079",
    isbn = "978-0-511-25227-3, 978-0-521-67227-6, 978-0-521-63303-1",
    publisher = "Cambridge University Press",
    series = "Cambridge Monographs on Mathematical Physics",
    month = "12",
    year = "2007"
}

@article{Rubakov:1983bz,
    author = "Rubakov, V. A. and Shaposhnikov, M. E.",
    title = "{Extra Space-Time Dimensions: Towards a Solution to the Cosmological Constant Problem}",
    reportNumber = "IC/83/11",
    doi = "10.1016/0370-2693(83)91254-6",
    journal = "Phys. Lett. B",
    volume = "125",
    pages = "139",
    year = "1983"
}

@article{Gogberashvili:1998vx,
    author = "Gogberashvili, Merab",
    title = "{Hierarchy problem in the shell universe model}",
    eprint = "hep-ph/9812296",
    archivePrefix = "arXiv",
    doi = "10.1142/S0218271802002992",
    journal = "Int. J. Mod. Phys. D",
    volume = "11",
    pages = "1635--1638",
    year = "2002"
}

@article{Gogberashvili:1998iu,
    author = "Gogberashvili, Merab",
    title = "{Our world as an expanding shell}",
    eprint = "hep-ph/9812365",
    archivePrefix = "arXiv",
    doi = "10.1209/epl/i2000-00162-1",
    journal = "EPL",
    volume = "49",
    pages = "396--399",
    year = "2000"
}

@article{Gogberashvili:1999tb,
    author = "Gogberashvili, Merab",
    title = "{Four dimensionality in noncompact Kaluza-Klein model}",
    eprint = "hep-ph/9904383",
    archivePrefix = "arXiv",
    doi = "10.1142/S021773239900208X",
    journal = "Mod. Phys. Lett. A",
    volume = "14",
    pages = "2025--2032",
    year = "1999"
}

@article{Randall:1999ee,
    author = "Randall, Lisa and Sundrum, Raman",
    title = "{A Large mass hierarchy from a small extra dimension}",
    eprint = "hep-ph/9905221",
    archivePrefix = "arXiv",
    reportNumber = "MIT-CTP-2860, PUPT-1860, BUHEP-99-9",
    doi = "10.1103/PhysRevLett.83.3370",
    journal = "Phys. Rev. Lett.",
    volume = "83",
    pages = "3370--3373",
    year = "1999"
}

@article{Randall:1999vf,
    author = "Randall, Lisa and Sundrum, Raman",
    title = "{An Alternative to compactification}",
    eprint = "hep-th/9906064",
    archivePrefix = "arXiv",
    reportNumber = "MIT-CTP-2874, PUPT-1867, BUHEP-99-13",
    doi = "10.1103/PhysRevLett.83.4690",
    journal = "Phys. Rev. Lett.",
    volume = "83",
    pages = "4690--4693",
    year = "1999"
}

@article{Appelquist:1983vs,
    author = "Appelquist, Thomas and Chodos, Alan",
    title = "{The Quantum Dynamics of Kaluza-Klein Theories}",
    reportNumber = "YTP 83-05",
    doi = "10.1103/PhysRevD.28.772",
    journal = "Phys. Rev. D",
    volume = "28",
    pages = "772",
    year = "1983"
}

@article{Garriga:2000jb,
    author = "Garriga, Jaume and Pujolas, Oriol and Tanaka, Takahiro",
    title = "{Radion effective potential in the brane world}",
    eprint = "hep-th/0004109",
    archivePrefix = "arXiv",
    reportNumber = "UAB-FT-486",
    doi = "10.1016/S0550-3213(01)00144-4",
    journal = "Nucl. Phys. B",
    volume = "605",
    pages = "192--214",
    year = "2001"
}

@article{Toms:2000bh,
    author = "Toms, David J.",
    title = "{Quantized bulk fields in the Randall-Sundrum compactification model}",
    eprint = "hep-th/0005189",
    archivePrefix = "arXiv",
    doi = "10.1016/S0370-2693(00)00618-3",
    journal = "Phys. Lett. B",
    volume = "484",
    pages = "149--153",
    year = "2000"
}

@article{Goldberger:2000dv,
    author = "Goldberger, Walter D. and Rothstein, Ira Z.",
    title = "{Quantum stabilization of compactified AdS(5)}",
    eprint = "hep-th/0007065",
    archivePrefix = "arXiv",
    reportNumber = "CALT-68-2286",
    doi = "10.1016/S0370-2693(00)01047-9",
    journal = "Phys. Lett. B",
    volume = "491",
    pages = "339--344",
    year = "2000"
}

@article{Brevik:2000vt,
    author = "Brevik, Iver H. and Milton, Kimball A. and Nojiri, Shin'ichi and Odintsov, Sergei D.",
    title = "{Quantum (in)stability of a brane world AdS(5) universe at nonzero temperature}",
    eprint = "hep-th/0010205",
    archivePrefix = "arXiv",
    reportNumber = "OKHEP-00-11",
    doi = "10.1016/S0550-3213(01)00026-8",
    journal = "Nucl. Phys. B",
    volume = "599",
    pages = "305--318",
    year = "2001"
}

@article{Greene:2007xu,
    author = "Greene, Brian R. and Levin, Janna",
    title = "{Dark Energy and Stabilization of Extra Dimensions}",
    eprint = "0707.1062",
    archivePrefix = "arXiv",
    primaryClass = "hep-th",
    doi = "10.1088/1126-6708/2007/11/096",
    journal = "JHEP",
    volume = "11",
    pages = "096",
    year = "2007"
}

@article{Kachru:2003aw,
    author = "Kachru, Shamit and Kallosh, Renata and Linde, Andrei D. and Trivedi, Sandip P.",
    title = "{De Sitter vacua in string theory}",
    eprint = "hep-th/0301240",
    archivePrefix = "arXiv",
    reportNumber = "SLAC-PUB-9630, SU-ITP-03-01, TIFR-TH-03-03",
    doi = "10.1103/PhysRevD.68.046005",
    journal = "Phys. Rev. D",
    volume = "68",
    pages = "046005",
    year = "2003"
}

@article{Balasubramanian:2005zx,
    author = "Balasubramanian, Vijay and Berglund, Per and Conlon, Joseph P. and Quevedo, Fernando",
    title = "{Systematics of moduli stabilisation in Calabi-Yau flux compactifications}",
    eprint = "hep-th/0502058",
    archivePrefix = "arXiv",
    reportNumber = "DAMTP-2005-10, UNH-05-01, UPR-1109-T",
    doi = "10.1088/1126-6708/2005/03/007",
    journal = "JHEP",
    volume = "03",
    pages = "007",
    year = "2005"
}

@article{Montero:2022prj,
    author = "Montero, Miguel and Vafa, Cumrun and Valenzuela, Irene",
    title = "{The dark dimension and the Swampland}",
    eprint = "2205.12293",
    archivePrefix = "arXiv",
    primaryClass = "hep-th",
    doi = "10.1007/JHEP02(2023)022",
    journal = "JHEP",
    volume = "02",
    pages = "022",
    year = "2023"
}

@article{Burgess:2023pnk,
    author = "Burgess, C. P. and Quevedo, F.",
    title = "{Perils of towers in the swamp: dark dimensions and the robustness of EFTs}",
    eprint = "2304.03902",
    archivePrefix = "arXiv",
    primaryClass = "hep-th",
    doi = "10.1007/JHEP09(2023)159",
    journal = "JHEP",
    volume = "09",
    pages = "159",
    year = "2023"
}

@article{Branchina:2023ogv,
    author = "Branchina, Carlo and Branchina, Vincenzo and Contino, Filippo and Pernace, Arcangelo",
    title = "{Does the cosmological constant really indicate the existence of a dark dimension?}",
    eprint = "2308.16548",
    archivePrefix = "arXiv",
    primaryClass = "hep-th",
    doi = "10.1142/S0219887824503055",
    journal = "Int. J. Geom. Meth. Mod. Phys.",
    volume = "22",
    number = "04",
    pages = "2450305",
    year = "2025"
}

@article{Anchordoqui:2023wkm,
    author = "Anchordoqui, Luis A. and Antoniadis, Ignatios and Cunat, Jules",
    title = "{Dark dimension and the standard model landscape}",
    eprint = "2306.16491",
    archivePrefix = "arXiv",
    primaryClass = "hep-ph",
    doi = "10.1103/PhysRevD.109.016028",
    journal = "Phys. Rev. D",
    volume = "109",
    number = "1",
    pages = "016028",
    year = "2024"
}

@article{Anchordoqui:2023laz,
    author = {Anchordoqui, Luis A. and Antoniadis, Ignatios and Lust, Dieter and L{\"u}st, Severin},
    title = "{On the cosmological constant, the KK mass scale, and the cut-off dependence in the dark dimension scenario}",
    eprint = "2309.09330",
    archivePrefix = "arXiv",
    primaryClass = "hep-th",
    reportNumber = "LMU-ASC 31/23, MPP-2023-191",
    doi = "10.1140/epjc/s10052-023-12206-2",
    journal = "Eur. Phys. J. C",
    volume = "83",
    number = "11",
    pages = "1016",
    year = "2023"
}

@article{Cui:2023wzo,
    author = "Cui, Chuanxin and Ning, Sirui",
    title = "{Casimir Energy Stabilization of Standard Model Landscape in Dark Dimension}",
    eprint = "2310.19592",
    archivePrefix = "arXiv",
    primaryClass = "hep-th",
    month = "10",
    year = "2023"
}

@article{Heckman:2024trz,
    author = "Heckman, Jonathan J. and Vafa, Cumrun and Weigand, Timo and Xu, Fengjun",
    title = "{Dark dimension and the grand unification of forces}",
    eprint = "2409.01405",
    archivePrefix = "arXiv",
    primaryClass = "hep-th",
    reportNumber = "ZMP-HH/24-19",
    doi = "10.1103/PhysRevD.111.046014",
    journal = "Phys. Rev. D",
    volume = "111",
    number = "4",
    pages = "046014",
    year = "2025"
}

@article{Anchordoqui:2025nmb,
    author = "Anchordoqui, Luis and Antoniadis, Ignatios and Lust, Dieter",
    title = "{Two Micron-Size Dark Dimensions}",
    eprint = "2501.11690",
    archivePrefix = "arXiv",
    primaryClass = "hep-th",
    reportNumber = "MPP-2025-5, LMU-ASC 02/25",
    month = "1",
    year = "2025"
}

@article{Banks:1999eg,
    author = "Banks, Tom and Dine, Michael and Nelson, Ann E.",
    title = "{Constraints on theories with large extra dimensions}",
    eprint = "hep-th/9903019",
    archivePrefix = "arXiv",
    reportNumber = "SCIPP-99-03, RU-98-53",
    doi = "10.1088/1126-6708/1999/06/014",
    journal = "JHEP",
    volume = "06",
    pages = "014",
    year = "1999"
}

@article{Cullen:1999hc,
    author = "Cullen, Schuyler and Perelstein, Maxim",
    title = "{SN1987A constraints on large compact dimensions}",
    eprint = "hep-ph/9903422",
    archivePrefix = "arXiv",
    reportNumber = "SLAC-PUB-8084, SU-ITP-99-15",
    doi = "10.1103/PhysRevLett.83.268",
    journal = "Phys. Rev. Lett.",
    volume = "83",
    pages = "268--271",
    year = "1999"
}

@article{Barger:1999jf,
    author = "Barger, Vernon D. and Han, Tao and Kao, C. and Zhang, Ren-Jie",
    title = "{Astrophysical constraints on large extra dimensions}",
    eprint = "hep-ph/9905474",
    archivePrefix = "arXiv",
    reportNumber = "MADPH-99-1118",
    doi = "10.1016/S0370-2693(99)00795-9",
    journal = "Phys. Lett. B",
    volume = "461",
    pages = "34--42",
    year = "1999"
}

@article{Hanhart:2000er,
    author = "Hanhart, Christoph and Phillips, Daniel R. and Reddy, Sanjay and Savage, Martin J.",
    title = "{Extra dimensions, SN1987a, and nucleon-nucleon scattering data}",
    eprint = "nucl-th/0007016",
    archivePrefix = "arXiv",
    reportNumber = "DOE-ER-40561-99-INT00, NT-UW-00-17, JLAB-THY-00-59, NT@UW-00-17",
    doi = "10.1016/S0550-3213(00)00667-2",
    journal = "Nucl. Phys. B",
    volume = "595",
    pages = "335--359",
    year = "2001"
}

@article{Hanhart:2001fx,
    author = "Hanhart, Christoph and Pons, Jose A. and Phillips, Daniel R. and Reddy, Sanjay",
    title = "{The Likelihood of GODs' existence: Improving the SN1987a constraint on the size of large compact dimensions}",
    eprint = "astro-ph/0102063",
    archivePrefix = "arXiv",
    doi = "10.1016/S0370-2693(01)00544-5",
    journal = "Phys. Lett. B",
    volume = "509",
    pages = "1--9",
    year = "2001"
}

@article{Hannestad:2003yd,
    author = "Hannestad, Steen and Raffelt, Georg G.",
    title = "{Supernova and neutron star limits on large extra dimensions reexamined}",
    eprint = "hep-ph/0304029",
    archivePrefix = "arXiv",
    reportNumber = "MPP-2003-19, MPP-2003-19",
    doi = "10.1103/PhysRevD.69.029901",
    journal = "Phys. Rev. D",
    volume = "67",
    pages = "125008",
    year = "2003",
    note = "[Erratum: Phys.Rev.D 69, 029901 (2004)]"
}

@article{Vafa:2024fpx,
    author = "Vafa, Cumrun",
    title = "{Swamplandish Unification of the Dark Sector}",
    eprint = "2402.00981",
    archivePrefix = "arXiv",
    primaryClass = "hep-ph",
    month = "2",
    year = "2024"
}

@book{Birrell:1982ix,
    author = "Birrell, N. D. and Davies, P. C. W.",
    title = "{Quantum Fields in Curved Space}",
    doi = "10.1017/CBO9780511622632",
    isbn = "978-0-511-62263-2, 978-0-521-27858-4",
    publisher = "Cambridge University Press",
    address = "Cambridge, UK",
    series = "Cambridge Monographs on Mathematical Physics",
    year = "1982"
}

@book{Buchbinder:1992rb,
    author = "Buchbinder, I. L. and Odintsov, S. D. and Shapiro, I. L.",
    title = "{Effective action in quantum gravity}",
    year = "1992"
}

@article{Stelle:1976gc,
    author = "Stelle, K. S.",
    title = "{Renormalization of Higher Derivative Quantum Gravity}",
    reportNumber = "PRINT-76-1059 (BRANDEIS)",
    doi = "10.1103/PhysRevD.16.953",
    journal = "Phys. Rev. D",
    volume = "16",
    pages = "953--969",
    year = "1977"
}

@article{Horowitz:1978fq,
    author = "Horowitz, G. T. and Wald, Robert M.",
    title = "{Dynamics of Einstein's Equation Modified by a Higher Order Derivative Term}",
    doi = "10.1103/PhysRevD.17.414",
    journal = "Phys. Rev. D",
    volume = "17",
    pages = "414--416",
    year = "1978"
}

@article{Horowitz:1980fj,
    author = "Horowitz, G. T.",
    title = "{SEMICLASSICAL RELATIVITY: THE WEAK FIELD LIMIT}",
    doi = "10.1103/PhysRevD.21.1445",
    journal = "Phys. Rev. D",
    volume = "21",
    pages = "1445--1461",
    year = "1980"
}

@article{Hartle:1981zt,
    author = "Hartle, J. B. and Horowitz, G. T.",
    title = "{Ground State Expectation Value of the Metric in the 1/$N$ or Semiclassical Approximation to Quantum Gravity}",
    doi = "10.1103/PhysRevD.24.257",
    journal = "Phys. Rev. D",
    volume = "24",
    pages = "257--274",
    year = "1981"
}

@article{RandjbarDaemi:1981wd,
    author = "Randjbar-Daemi, S.",
    title = "{Stability of the Minkowski Vacuum in the Renormalized Semiclassical Theory of Gravity}",
    reportNumber = "ICTP/80-81/11",
    doi = "10.1088/0305-4470/14/7/001",
    journal = "J. Phys. A",
    volume = "14",
    pages = "L229",
    year = "1981"
}

@article{Jordan:1987wd,
    author = "Jordan, R. D.",
    title = "{Stability of Flat Space-time in Quantum Gravity}",
    doi = "10.1103/PhysRevD.36.3593",
    journal = "Phys. Rev. D",
    volume = "36",
    pages = "3593--3603",
    year = "1987"
}

@article{Suen:1988uf,
    author = "Suen, Wai-Mo",
    title = "{The Stability of the Semiclassical Einstein Equation}",
    reportNumber = "PRINT-88-0861 (WASH.U.,ST.LOUIS)",
    doi = "10.1103/PhysRevD.40.315",
    journal = "Phys. Rev. D",
    volume = "40",
    pages = "315",
    year = "1989"
}

@article{Suen:1989bg,
    author = "Suen, W. M.",
    title = "{Minkowski Space-time Is Unstable in Semiclassical Gravity}",
    doi = "10.1103/PhysRevLett.62.2217",
    journal = "Phys. Rev. Lett.",
    volume = "62",
    pages = "2217--2220",
    year = "1989"
}

@article{Anderson:2002fk,
    author = "Anderson, Paul R. and Molina-Paris, Carmen and Mottola, Emil",
    title = "{Linear response, validity of semiclassical gravity, and the stability of flat space}",
    eprint = "gr-qc/0209075",
    archivePrefix = "arXiv",
    reportNumber = "LA-UR-02-5809",
    doi = "10.1103/PhysRevD.67.024026",
    journal = "Phys. Rev. D",
    volume = "67",
    pages = "024026",
    year = "2003"
}

@article{Matsui:2019tlf,
    author = "Matsui, Hiroki and Watamura, Naoki",
    title = "{Quantum Spacetime Instability and Breakdown of Semiclassical Gravity}",
    eprint = "1910.02186",
    archivePrefix = "arXiv",
    primaryClass = "gr-qc",
    reportNumber = "TU-1092",
    doi = "10.1103/PhysRevD.101.025014",
    journal = "Phys. Rev. D",
    volume = "101",
    number = "2",
    pages = "025014",
    year = "2020"
}

@article{Simon:1991bm,
    author = "Simon, Jonathan Z.",
    title = "{No Starobinsky inflation from selfconsistent semiclassical gravity}",
    reportNumber = "WISC-MILW-91-TH-11",
    doi = "10.1103/PhysRevD.45.1953",
    journal = "Phys. Rev. D",
    volume = "45",
    pages = "1953--1960",
    year = "1992"
}

@article{Parker:1993dk,
    author = "Parker, Leonard and Simon, Jonathan Z.",
    title = "{Einstein equation with quantum corrections reduced to second order}",
    eprint = "gr-qc/9211002",
    archivePrefix = "arXiv",
    reportNumber = "WISC-MILW-92-TH-14",
    doi = "10.1103/PhysRevD.47.1339",
    journal = "Phys. Rev. D",
    volume = "47",
    pages = "1339--1355",
    year = "1993"
}

@article{Arkani-Hamed:2007ryu,
    author = "Arkani-Hamed, Nima and Dubovsky, Sergei and Nicolis, Alberto and Villadoro, Giovanni",
    title = "{Quantum Horizons of the Standard Model Landscape}",
    eprint = "hep-th/0703067",
    archivePrefix = "arXiv",
    doi = "10.1088/1126-6708/2007/06/078",
    journal = "JHEP",
    volume = "06",
    pages = "078",
    year = "2007"
}

@article{Yu:1999pq,
    author = "Yu, Hong-wei and Ford, L. H.",
    title = "{Light cone fluctuations in flat space-times with nontrivial topology}",
    eprint = "gr-qc/9904082",
    archivePrefix = "arXiv",
    doi = "10.1103/PhysRevD.60.084023",
    journal = "Phys. Rev. D",
    volume = "60",
    pages = "084023",
    year = "1999"
}

@article{Yu:2000yf,
    author = "Yu, Hong-wei and Ford, L. H.",
    title = "{Quantum light cone fluctuations in theories with extra dimensions}",
    eprint = "gr-qc/0004063",
    archivePrefix = "arXiv",
    month = "4",
    year = "2000"
}

@article{Copeland:2006wr,
    author = "Copeland, Edmund J. and Sami, M. and Tsujikawa, Shinji",
    title = "{Dynamics of dark energy}",
    eprint = "hep-th/0603057",
    archivePrefix = "arXiv",
    doi = "10.1142/S021827180600942X",
    journal = "Int. J. Mod. Phys. D",
    volume = "15",
    pages = "1753--1936",
    year = "2006"
}

@article{DESI:2024mwx,
    author = "Adame, A. G. and others",
    collaboration = "DESI",
    title = "{DESI 2024 VI: cosmological constraints from the measurements of baryon acoustic oscillations}",
    eprint = "2404.03002",
    archivePrefix = "arXiv",
    primaryClass = "astro-ph.CO",
    reportNumber = "FERMILAB-PUB-24-0154-PPD",
    doi = "10.1088/1475-7516/2025/02/021",
    journal = "JCAP",
    volume = "02",
    pages = "021",
    year = "2025"
}

@article{DESI:2025zgx,
    author = "Abdul Karim, M. and others",
    collaboration = "DESI",
    title = "{DESI DR2 results. II. Measurements of baryon acoustic oscillations and cosmological constraints}",
    eprint = "2503.14738",
    archivePrefix = "arXiv",
    primaryClass = "astro-ph.CO",
    reportNumber = "FERMILAB-PUB-25-0169-PPD",
    doi = "10.1103/tr6y-kpc6",
    journal = "Phys. Rev. D",
    volume = "112",
    number = "8",
    pages = "083515",
    year = "2025"
}

@article{Carroll:2001ih,
    author = "Carroll, Sean M. and Geddes, James and Hoffman, Mark B. and Wald, Robert M.",
    title = "{Classical stabilization of homogeneous extra dimensions}",
    eprint = "hep-th/0110149",
    archivePrefix = "arXiv",
    reportNumber = "EFI-2001-38",
    doi = "10.1103/PhysRevD.66.024036",
    journal = "Phys. Rev. D",
    volume = "66",
    pages = "024036",
    year = "2002"
}

@article{DeFelice:2010aj,
    author = "De Felice, Antonio and Tsujikawa, Shinji",
    title = "{f(R) theories}",
    eprint = "1002.4928",
    archivePrefix = "arXiv",
    primaryClass = "gr-qc",
    doi = "10.12942/lrr-2010-3",
    journal = "Living Rev. Rel.",
    volume = "13",
    pages = "3",
    year = "2010"
}

@article{Deruelle:2010ht,
    author = "Deruelle, Nathalie and Sasaki, Misao",
    editor = "Odintsov, Sergey D. and S{\'a}ez-G{\'o}mez, Diego and Xamb{\'o}-Descamps, Sebastian",
    title = "{Conformal equivalence in classical gravity: the example of 'Veiled' General Relativity}",
    eprint = "1007.3563",
    archivePrefix = "arXiv",
    primaryClass = "gr-qc",
    reportNumber = "YITP-10-61",
    doi = "10.1007/978-3-642-19760-4_23",
    journal = "Springer Proc. Phys.",
    volume = "137",
    pages = "247--260",
    year = "2011"
}

@article{Chiba:2013mha,
    author = "Chiba, Takeshi and Yamaguchi, Masahide",
    title = "{Conformal-Frame (In)dependence of Cosmological Observations in Scalar-Tensor Theory}",
    eprint = "1308.1142",
    archivePrefix = "arXiv",
    primaryClass = "gr-qc",
    doi = "10.1088/1475-7516/2013/10/040",
    journal = "JCAP",
    volume = "10",
    pages = "040",
    year = "2013"
}

@article{Kesden:2006zb,
    author = "Kesden, Michael and Kamionkowski, Marc",
    title = "{Galilean Equivalence for Galactic Dark Matter}",
    eprint = "astro-ph/0606566",
    archivePrefix = "arXiv",
    doi = "10.1103/PhysRevLett.97.131303",
    journal = "Phys. Rev. Lett.",
    volume = "97",
    pages = "131303",
    year = "2006"
}

@article{Bedroya:2025fwh,
    author = "Bedroya, Alek and Obied, Georges and Vafa, Cumrun and Wu, David H.",
    title = "{Evolving Dark Sector and the Dark Dimension Scenario}",
    eprint = "2507.03090",
    archivePrefix = "arXiv",
    primaryClass = "astro-ph.CO",
    month = "7",
    year = "2025"
}

@article{Das:2005yj,
    author = "Das, Subinoy and Corasaniti, Pier Stefano and Khoury, Justin",
    title = "{Super-acceleration as signature of dark sector interaction}",
    eprint = "astro-ph/0510628",
    archivePrefix = "arXiv",
    doi = "10.1103/PhysRevD.73.083509",
    journal = "Phys. Rev. D",
    volume = "73",
    pages = "083509",
    year = "2006"
}

@article{DESI:2024hhd,
    author = "Adame, A. G. and others",
    collaboration = "DESI",
    title = "{DESI 2024 VII: cosmological constraints from the full-shape modeling of clustering measurements}",
    eprint = "2411.12022",
    archivePrefix = "arXiv",
    primaryClass = "astro-ph.CO",
    reportNumber = "FERMILAB-PUB-24-0854-PPD",
    doi = "10.1088/1475-7516/2025/07/028",
    journal = "JCAP",
    volume = "07",
    pages = "028",
    year = "2025"
}

@article{DESI:2025fii,
    author = "Lodha, K. and others",
    collaboration = "DESI",
    title = "{Extended dark energy analysis using DESI DR2 BAO measurements}",
    eprint = "2503.14743",
    archivePrefix = "arXiv",
    primaryClass = "astro-ph.CO",
    reportNumber = "FERMILAB-PUB-25-0164-PPD",
    doi = "10.1103/w4c6-1r5j",
    journal = "Phys. Rev. D",
    volume = "112",
    number = "8",
    pages = "083511",
    year = "2025"
}

@article{Koivisto:2013fta,
    author = "Koivisto, Tomi and Wills, Danielle and Zavala, Ivonne",
    title = "{Dark D-brane Cosmology}",
    eprint = "1312.2597",
    archivePrefix = "arXiv",
    primaryClass = "hep-th",
    reportNumber = "NORDITA-2013-99, DCPT-13-55",
    doi = "10.1088/1475-7516/2014/06/036",
    journal = "JCAP",
    volume = "06",
    pages = "036",
    year = "2014"
}

@article{vandeBruck:2015ida,
    author = "van de Bruck, C. and Morrice, J.",
    title = "{Disformal couplings and the dark sector of the universe}",
    eprint = "1501.03073",
    archivePrefix = "arXiv",
    primaryClass = "gr-qc",
    doi = "10.1088/1475-7516/2015/04/036",
    journal = "JCAP",
    volume = "04",
    pages = "036",
    year = "2015"
}

@article{Hardy:2025ajb,
    author = "Hardy, Edward and Sokolov, Anton and Stubbs, Henry",
    title = "{Stellar cooling limits on KK gravitons and dark dimensions}",
    eprint = "2510.18975",
    archivePrefix = "arXiv",
    primaryClass = "hep-ph",
    month = "10",
    year = "2025"
}

@article{Ooguri:2006in,
    author = "Ooguri, Hirosi and Vafa, Cumrun",
    title = "{On the Geometry of the String Landscape and the Swampland}",
    eprint = "hep-th/0605264",
    archivePrefix = "arXiv",
    reportNumber = "CALT-68-2600, HUTP-06-A017",
    doi = "10.1016/j.nuclphysb.2006.10.033",
    journal = "Nucl. Phys. B",
    volume = "766",
    pages = "21--33",
    year = "2007"
}

@article{Lust:2019zwm,
    author = {L{\"u}st, Dieter and Palti, Eran and Vafa, Cumrun},
    title = "{AdS and the Swampland}",
    eprint = "1906.05225",
    archivePrefix = "arXiv",
    primaryClass = "hep-th",
    doi = "10.1016/j.physletb.2019.134867",
    journal = "Phys. Lett. B",
    volume = "797",
    pages = "134867",
    year = "2019"
}

@article{Chevallier:2000qy,
    author = "Chevallier, Michel and Polarski, David",
    title = "{Accelerating universes with scaling dark matter}",
    eprint = "gr-qc/0009008",
    archivePrefix = "arXiv",
    doi = "10.1142/S0218271801000822",
    journal = "Int. J. Mod. Phys. D",
    volume = "10",
    pages = "213--224",
    year = "2001"
}

@article{Linder:2002et,
    author = "Linder, Eric V.",
    title = "{Exploring the expansion history of the universe}",
    eprint = "astro-ph/0208512",
    archivePrefix = "arXiv",
    doi = "10.1103/PhysRevLett.90.091301",
    journal = "Phys. Rev. Lett.",
    volume = "90",
    pages = "091301",
    year = "2003"
}

@article{Lee:2019wij,
    author = "Lee, Seung-Joo and Lerche, Wolfgang and Weigand, Timo",
    title = "{Emergent strings from infinite distance limits}",
    eprint = "1910.01135",
    archivePrefix = "arXiv",
    primaryClass = "hep-th",
    reportNumber = "CERN-TH-2019-159",
    doi = "10.1007/JHEP02(2022)190",
    journal = "JHEP",
    volume = "02",
    pages = "190",
    year = "2022"
}

@article{Lee:2019xtm,
    author = "Lee, Seung-Joo and Lerche, Wolfgang and Weigand, Timo",
    title = "{Emergent strings, duality and weak coupling limits for two-form fields}",
    eprint = "1904.06344",
    archivePrefix = "arXiv",
    primaryClass = "hep-th",
    reportNumber = "CERN-TH-2019-044",
    doi = "10.1007/JHEP02(2022)096",
    journal = "JHEP",
    volume = "02",
    pages = "096",
    year = "2022"
}

@article{Huey:2004qv,
    author = "Huey, Greg and Wandelt, Benjamin D.",
    title = "{Interacting quintessence. The Coincidence problem and cosmic acceleration}",
    eprint = "astro-ph/0407196",
    archivePrefix = "arXiv",
    doi = "10.1103/PhysRevD.74.023519",
    journal = "Phys. Rev. D",
    volume = "74",
    pages = "023519",
    year = "2006"
}

@article{Eisenstein:1997ik,
    author = "Eisenstein, Daniel J. and Hu, Wayne",
    title = "{Baryonic features in the matter transfer function}",
    eprint = "astro-ph/9709112",
    archivePrefix = "arXiv",
    reportNumber = "IASSNS-AST-97-51",
    doi = "10.1086/305424",
    journal = "Astrophys. J.",
    volume = "496",
    pages = "605",
    year = "1998"
}

@article{SDSS:2005xqv,
    author = "Eisenstein, Daniel J. and others",
    collaboration = "SDSS",
    title = "{Detection of the Baryon Acoustic Peak in the Large-Scale Correlation Function of SDSS Luminous Red Galaxies}",
    eprint = "astro-ph/0501171",
    archivePrefix = "arXiv",
    reportNumber = "FERMILAB-PUB-05-057-A-CD",
    doi = "10.1086/466512",
    journal = "Astrophys. J.",
    volume = "633",
    pages = "560--574",
    year = "2005"
}

@article{Bassett:2009mm,
    author = "Bassett, Bruce A. and Hlozek, Renee",
    title = "{Baryon Acoustic Oscillations}",
    eprint = "0910.5224",
    archivePrefix = "arXiv",
    primaryClass = "astro-ph.CO",
    month = "10",
    year = "2009"
}

@article{Lee:2020zjt,
    author = "Lee, J. G. and Adelberger, E. G. and Cook, T. S. and Fleischer, S. M. and Heckel, B. R.",
    title = "{New Test of the Gravitational $1/r^2$ Law at Separations down to 52 $\mu$m}",
    eprint = "2002.11761",
    archivePrefix = "arXiv",
    primaryClass = "hep-ex",
    doi = "10.1103/PhysRevLett.124.101101",
    journal = "Phys. Rev. Lett.",
    volume = "124",
    number = "10",
    pages = "101101",
    year = "2020"
}
\bibliographystyle{JHEP}

\end{document}